\documentclass[aps,twocolumn,english, superscriptaddress,eqsecnum,showpacs,longbibliography]{revtex4-1}
\usepackage{amssymb}
\usepackage{amsmath}
\usepackage{graphicx}
\usepackage{xcolor}
\usepackage{epsfig}
\usepackage[colorlinks,citecolor=blue,linkcolor=red,urlcolor=blue]{hyperref}
\usepackage[all]{hypcap}
\usepackage{braket}
\usepackage[normalem]{ulem}

\usepackage{babel}
\usepackage{xcolor}
\usepackage{wasysym}
\usepackage{hhline}
\usepackage{ulem}
\usepackage{booktabs}

\newcommand{\be}{\begin{equation}}
\newcommand{\ee}{\end{equation}}
\newcommand{\bea}{\begin{eqnarray}}
\newcommand{\eea}{\end{eqnarray}}

\usepackage[all]{hypcap} 

\begin{document}

\title{Dynamical properties of the $S=\frac{1}{2}$ random Heisenberg chain}

\author{Yu-Rong Shu}
\affiliation{State Key Laboratory of Optoelectronic Materials and Technologies, School of Physics, Sun Yat-Sen University, Guangzhou 510275, China}
\affiliation{Department of Physics, Boston University, 590 Commonwealth Avenue, Boston, Massachusetts 02215, USA}

\author{Maxime Dupont}
\affiliation{Department of Physics, Boston University, 590 Commonwealth Avenue, Boston, Massachusetts 02215, USA}
\affiliation{Laboratoire de Physique Th\'eorique, IRSAMC, Universit\'e de Toulouse, CNRS, F-31062, Toulouse, France}

\author{Dao-Xin Yao}
\email{yaodaox@mail.sysu.edu.cn}
\affiliation{State Key Laboratory of Optoelectronic Materials and Technologies, School of Physics, Sun Yat-Sen University, Guangzhou 510275, China}

\author{Sylvain Capponi}
\email{capponi@irsamc.ups-tlse.fr}
\affiliation{Department of Physics, Boston University, 590 Commonwealth Avenue, Boston, Massachusetts 02215, USA}
\affiliation{Laboratoire de Physique Th\'eorique, IRSAMC, Universit\'e de Toulouse, CNRS, F-31062, Toulouse, France}

\author{Anders W. Sandvik}
\email{sandvik@bu.edu}
\affiliation{Department of Physics, Boston University, 590 Commonwealth Avenue, Boston, Massachusetts 02215, USA}

\date{\today}

\begin{abstract}
We study dynamical properties at finite temperature ($T$) of Heisenberg spin chains with random antiferrormagnetic exchange couplings, which realize the
random singlet phase in the low-energy limit, using three complementary numerical methods: exact diagonalization, matrix-product-state algorithms, and
stochastic analytic continuation of quantum Monte Carlo results in imaginary time. Specifically, we investigate the dynamic spin structure factor
$S(q,\omega)$ and its $\omega\rightarrow 0$ limit, which are closely related to inelastic neutron scattering and nuclear magnetic resonance
(NMR) experiments (through the spin-lattice relaxation rate $1/T_1$). Our study reveals a continuous narrow band of low-energy excitations in
$S(q,\omega)$, extending throughout the $q$-space, instead of being restricted to $q\approx 0$ and $q\approx \pi$ as found in the uniform system.
Close to $q=\pi$, the scaling properties of these excitations are well captured by the random-singlet theory, but disagreements also exist with some
aspects of the predicted $q$-dependence further away from $q=\pi$. Furthermore we also find spin diffusion effects close to $q=0$ that are not contained
within the random-singlet theory but give non-negligible contributions to the mean $1/T_1$. To compare with NMR experiments, we consider the distribution
of the local relaxation rates $1/T_{1}$. We show that the local $1/T_1$ values are broadly distributed,
approximately according to a stretched exponential. The mean $1/T_{1}$ first decreases with $T$, but below a crossover temperature it starts to increase
and likely diverges in the limit of a small nuclear resonance frequency $\omega_0$. Although a similar divergent behavior has been predicted and
experimentally observed for the static uniform susceptibility, this divergent behavior of the mean $1/T_1$ has never been experimentally observed.
Indeed, we show that the divergence of the mean $1/T_1$ is due to rare events in the disordered chains and is concealed in experiments, where the typical
$1/T_1$ value is accessed.

\end{abstract}

\maketitle

\section{Introduction}\label{sec:intro}

The apparent simplicity of one-dimensional ($1$D) spin chains provide an ideal testing ground for theories, experiments, and numerical simulations
in quantum magnetism. Disorder in such a system can drastically change its properties, or it may remain insensitive to weak disorder, thus qualifying
the disorder as irrelevant. Controlling the disorder strength can thus drive a quantum phase transition between distinct phases. For instance, the
spin$-1/2$ antiferromagnetic (AF) Heisenberg chain with uniform nearest-neighbor exchange couplings $J$ can be effectively described in the low-energy sector as
a Tomonaga-Luttinger liquid (TLL) \cite{giamarchi2003}, and it can also be solved exactly using the Bethe ansatz. However, the presence of any amount of
disorder in the exchange couplings will bring the system into the completely different random-singlet (RS) phase~\cite{dasgupta1980,fisher1994,igloi2005}.
This state can be studied by means of the strong-disorder renormalization group (SDRG) scheme \cite{igloi2005}, where singlets are successively formed between
the two spins with the strongest exchange coupling at each stage, thereby decimating this spin pair from the system but yielding a new effective interaction
between the spins previously coupled to it. The resulting asymptotic state can be nicely pictured as a collection of spins paired up into singlets spanning
arbitrary distances (with the RS theory yielding an asymptotically exact distribution of the distances). Though some of the most important properties of the
ground state, and also some dynamic and thermodynamic properties, can be understood based on the SDRG framework, this simple scheme cannot address all relevant
questions, especially as concerns dynamical properties. It is therefore useful to apply a wider range of analytical and numerical tools to these systems,
especially in light of the fact that there are also good experimental realizations of the random Heisenberg chain. We will here apply several numerical
methods to the random-exchange chain and study the temperature dependence of its dynamic spin structure factor, $S(q,\omega)$, in the full wavenumber ($q$)
and frequency ($\omega$) space. This quantity is important in the context of inelastic neutron scattering (INS) and nuclear magnetic resonance (NMR) experiments,
and also represents the most basic spectral function of interest in theoretical studies of uniform and random quantum spin models.

The radical change in the ground state due to disorder naturally affects the low-energy properties of the $S=1/2$ chain, where at low temperatures one can
expect that excitations of the weak singlets---those that have been decimated at the latter stages in the SDRG procedure---will give rise to low-energy
excitations not present in the uniform system. One well known consequence of this enhanced density of low-energy excitations is that the magnetic
susceptibility $\chi(T)$, which in the pure chain takes a finite value as $T \to 0$, diverges as $1/T$ in the RS phase \cite{dasgupta1980,fisher1994}
(and in both systems there are logarithmic corrections \cite{griffiths1964,eggert1994}). This behavior was first observed
experimentally in a class of quasi-1D Bechgaard salts \cite{bulaevskii1972,azevedo1977,sanny1980,bozler1980,tippie1981,theodorou1976,theodorou1977_1,theodorou1977_2},
and more recently in other quasi-2D compounds such as BaCu$_2$(Si$_{0.5}$Ge$_{0.5}$)$_2$2O$_7$ \cite{masuda2004} and BaCu$_2$SiGeO$_7$ \cite{shiroka2013}. The
specific heat, which for a TLL is linear in temperature, $C(T)\sim T$
\cite{bonner1964,affleck1986}, instead approaches a non-zero
constant as $T \to 0$ in the RS phase. Though the mean spin-spin correlation function $\langle\mathbf{S}_i\cdot\mathbf{S}_{i+r}\rangle$ remains
algebraically decaying in the RS phase, it changes from the form $\sim (-1)^rr^{-\eta}$ with $\eta=1$ for the clean chain to $\eta=2$ in the presence of disorder
(and in both cases there is a multiplicative logarithmic correction to the dominant power law \cite{affleck1989,singh1989,giamarchi1989,shu2016}).

Dynamical quantities are expected to undergo drastic changes as well and there are some fairly detailed RS theory predictions
for the $q$ and $\omega$ dependence on $S(q,\omega)$ for $q$ close to $\pi$ and $0$ \cite{damle2000,motrunich2001}. The full $S(q,\omega)$ can be
mapped out in INS experiments while in NMR experiments,
the inverse spin-lattice relaxation time $1/T_1$ is directly related in the simplest case to the $q$-integrated dynamic structure factor $S_0(\omega)$ at
the resonance frequency $\omega=\omega_0$, where normally $\omega_0 \ll J$ and one can consider the limit $\omega_0 \to 0$ (though some times the dependence on
$\omega_0$ can still be detected and gives additional information on the excitation spectrum). This local quantity $1/T_{1}$ diverges logarithmically at low $T$
for the pure chain~\cite{sachdev1994,sandvik1995,takigawa1997,barzykin2000,barzykin2001,coira2016,dupont2016} but with strong disorder, it was pointed out to
distributed according to a broad stretched exponential among disorder realizations with the mean value approaching zero as $T\rightarrow 0$ in both
experimental~\cite{shiroka2011} and numerical~\cite{herbrych2013} observations.
Furthermore, the presence of disorder brings a singular ($\delta$ peak) contribution at $\omega=0$ to $S(q,\omega)$~\cite{herbrych2013}, which
is not present in the uniformly coupled system. This $\delta$ peak at $\omega=0$ could be probed using INS experiments but is not relevant for NMR experiments performed
at small but nonzero resonance frequency and therefore should be excluded from the contributions to $1/T_{1}$.

Although dynamical observables such as $S(q,\omega)$ are more challenging to compute, numerical studies remain essential. The only truly unbiased numerical method
for dynamics is exact diagonalization (ED), which is limited to small system sizes but still provides important insights and benchmark results for testing other
methods. Density matrix renormalization group (DMRG) techniques~\cite{white1992,white1993}, or more generally the Matrix product state (MPS)
formalism~\cite{schollwock2011}, have proved to be efficient in dealing with large $1$D quantum systems. However, the time evolution of a state produces a rapid
growth of entanglement entropy~\cite{laflorencie2016}, and this causes convergence problems since the method relies on low-entangled states through the area law.
Thus, this method is limited to accessing only
short and intermediate times, which makes it difficult to resolve the important low-frequency behavior. Quantum Monte Carlo (QMC) simulations are not
limited by dimension nor system size but cannot access real-time dynamics directly owing to the ``sign problem''. Instead, one can compute imaginary-time correlations
and then, {\it a posteriori}, analytically continue from the imaginary axis to the real axis. This procedure is difficult due to the limited
information contained in the correlation functions in the presence of statistical sampling errors, resulting in a wide range of possible solutions
when transforming to real frequencies. To exclude unlikely solutions with large (noisy) frequency variations, the analytic continuation has to involve
some regularization mechanism. The so far most widely used approach is the maximum entropy (ME) method \cite{gull1984,silver1990,gubernatis1991}, which has
been very useful in constraining the solution by favoring a large entropy of the spectrum relative to a default model (thus giving a smoothly varying
spectrum). However, the ME method is often overly biased towards high-entropy spectra and leads to excessive broadening and other distortions of sharp spectral
features \cite{sandvik1998}. An alternative method, stochastic analytic continuation (SAC), also called the average-spectrum method, has been
developed~\cite{sandvik1998,beach2004,syljuasen2008,fuchs2010,sandvik2016} in which a suitably parametrized spectral function is sampled using Monte Carlo
methods. Here the entropy is intrinsic in the space of accessible spectral functions and the averaging has a regularizing effect similar to the entropy prior
in the ME method. However, the frequency resolution is often better and one can also control excessive entropic broadening by imposing and optimizing
constraints \cite{sandvik2016,shao2017}. Here we adopt the SAC method with the parametrization recently introduced and applied
in Refs.~\cite{qin2017,shao2017}.

An important aspect of this work is to compare the different approaches for computing $S(q,\omega)$ of the random Heisenberg chain, and establish the temperature
regimes in which reliable results can be obtained. By combining the results of the different methods in their respective optimal regimes, we are able to
reach a rather complete picture of the evolution of the spectral weight in frequency and momentum space as the temperature is lowered and the system gradually
approaches the RS fixed point. Our results confirm some aspects of the predictions from RS theory \cite{damle2000,motrunich2001}, in particular for $q$ close to
$\pi$, but also indicate its limitations in capturing the behavior of $S(q,\omega)$ away from $q=\pi$ and the significant spin-diffusion contributions around $q=0$.
In the meantime, we focus on the $q$-dependence of the spin-lattice relaxation rate and the distribution of the $1/T_{1}$ values from different disorder realizations.
We find distinction between the mean and typical values of the distribution, which discloses the discrepancy between the divergence of $1/T_{1}$ as $T\rightarrow 0$ in our
results and vanishing behavior found previously~\cite{shiroka2011,herbrych2013}.

The rest of the paper is organized as follows: In Sec.~\ref{sec:model_def} we introduce the properties of the model and the observables measured. We also discuss the
origin of the singular $\omega=0$ contribution to $S(q,\omega)$. In Sec.~\ref{sec:numerical_tech} we compare the result of $S(q,\omega)$ obtained using the three
different numerical approaches(ED, MPS and QMC-SAC). We present our results in Sec.~\ref{sec:results}. In Sec.~\ref{sec:conclusion} we summarize our conclusions
and discuss implications and open issues.

\section{Model and observables}\label{sec:model_def}

\subsection{Model}
We consider the $\mathrm{SU}(2)$ symmetric $S=1/2$ Heisenberg chain with random first-neighbor couplings, described by the Hamiltonian
\be
    \mathcal{H} = \sum_{i=1}^L J_{i}\mathbf{S}_{i}\cdot\mathbf{S}_{i+1},
    \label{eq:hamiltonian}
\ee
where the couplings $J_i$ are drawn from one out of several distributions to be specified further below. Both open and periodic boundary
conditions (OBCs and PBCs, respectively) are considered, depending on the numerical method.

Before discussing the disorder distributions and physical quantities we will study, it is useful to review the salient features of the
model that follow from its treatment with the SDRG method. This method amounts to decimation of spin pairs as discussed in Sec.~\ref{sec:intro},
and at each step the energy scale $\Omega$ is reduced. The meaning of $\Omega$ is that all remaining effective couplings (i.e., those
that will be generated during the decimation of the remaining spins) are less than $\Omega$. The distribution of effective couplings $J$
is given by~\cite{fisher1994}
\be
    P\left(J,\Omega\right) = \frac{-1}{\Omega\ln\Omega}\left(\frac{\Omega}{J}\right)^{1+1/\ln\Omega}\Theta\left(\Omega-J\right),
    \label{eq:rsdist}
\ee
where $\Theta(x)$ is the Heaviside step function and the initial energy scale $\Omega_0$ of the SDRG decimation procedure has been set to unity (or
else $\Omega/\Omega_0$ should be used under the logarithms). Here it is also assumed that the system size is very large (strictly speaking infinite).
Then, at the RS fixed point, the energy scale $\Omega$ vanishes and the coupling distribution becomes singular. While the energy scale $\Omega$ is concretely
defined in the context of a running decimation procedure, it also has direct physical interpretations as an energy cutoff in various situations, e.g.,
finite temperature can roughly be captured by setting the scale at $\Omega=T$.

In the Heisenberg chain, any amount of disorder in the exchange couplings will make the system eventually flow towards the RS fixed point~\cite{fisher1994},
though for a system with weak disorder it takes a large system size and low temperature to observe the crossover from the clean-chain behavior to the ultimate
RS properties; this crossover is also understood quantitatively \cite{laflorencie2004}. Many of works have been devoted to unbiased numerical studies
with various numerical methods, e.g., QMC simulations and the DMRG technique, to test the many predictions based on the SDRG
scheme for many different models. In the case of the Heisenberg chain \cite{hikihara1999,laflorencie2004,hoyos2007,shu2016}, these works have confirmed various
power-law behaviors associated with the RS phase, but also have made it clear that non-asymptotic corrections can partially mask the RS physics when considering
system sizes and temperatures that can be reached in practice. In addition, multiplicative logarithmic corrections have also been found \cite{shu2016}.

\begin{figure}[htbp]
    \includegraphics[width=\columnwidth,clip]{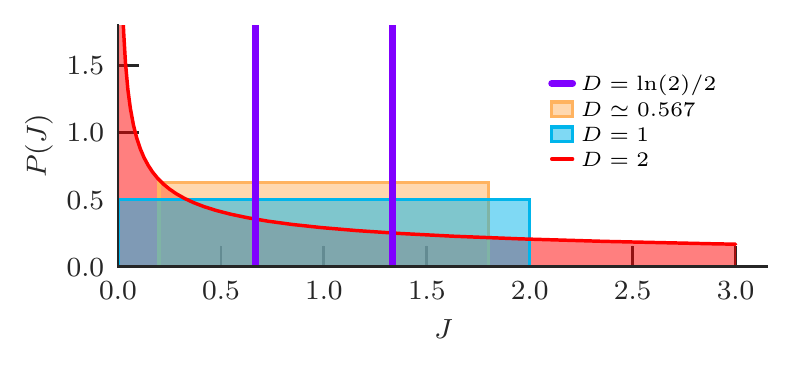}
\vskip-3mm
    \caption{Sketch of the four different disorder distributions of the random exchange couplings considered in this work.
      They are defined by Eq.~\eqref{eq:dist}~(except the bimodal distribution) and referred to by the disorder strength
      $D$, Eq.~\eqref{eq:dis_strength}. The mean couplings $\bar{J}=1$ in all cases.}
    \label{fig:PJ_sketch}
\end{figure}

In the present work, we use several different disorder distributions of the random exchange couplings, given by the distribution
\be
    P(J) = \frac{A}{d}J^{-1+1/d},\quad J_\mathrm{min} < J < J_\mathrm{max},
    \label{eq:dist}
\ee
in which the prefactor,
\be
A=1/(J_\mathrm{max}^{1/d}-J_\mathrm{min}^{1/d}),
\ee
is determined by the normalization of $P(J)$ and $d$ is a convenient parameter controlling the shape of the distribution. We quantify
the ``disorder strength'' $D$ by the variance of the distribution of the logarithm of the couplings~\cite{igloi2005};
\be
    D^{2}=\overline{(\ln J)^{2}}-\overline{\ln J}\,^{2}.
    \label{eq:dis_strength}
\ee
For $d=1$, Eq.~\eqref{eq:dist} generates random exchange couplings uniformly drawn from the box $(J_\mathrm{min},J_\mathrm{max})$, while when $d>1$ the
distribution has a power law form. In the limit of $d\rightarrow \infty$ and $J_\mathrm{min}=0$, Eq.~\eqref{eq:dist} corresponds to the singular distribution
at the RS fixed-point. Both the cases $d=1$ and $d=2$ with $J_\mathrm{min}=0$ are studied in the present work. One would expect that the asymptotic RS
behavior is better manifested (with less non-asymptotic contributions) if $d$ is large, but numerically, with the relatively small system sizes accessible
in practice, it is easier to compute proper disorder averages if $d$ is smaller. The values chosen here reflect a practical compromise in this regard.
We also consider the case $J_\mathrm{min}=0.2$ and $d=1$, to compare with previous results for the dynamic structure factor obtained with this distribution
\cite{herbrych2013}. The values of $J_\mathrm{max}$ are chosen by imposing that the average of all random couplings $\bar{J}$ equals $1$, to ensure that
the overall energy scale of the Hamiltonian, Eq.~\eqref{eq:hamiltonian}, equals unity in all cases. We finally also consider the bimodal distribution where
$J_i$ takes the values $2/3$ and $4/3$ with equal probability, in order to compare with experiments where this distribution has been proposed \cite{shiroka2011}.
Figure~\ref{fig:PJ_sketch} summarizes all the different disorder distributions considered in our work.

\subsection{Dynamic Structure Factor}\label{sec:sing_reg_contrib}

For the isotropic random Heisenberg chain we considered, the finite temperature dynamic structure factor is defined as follows in the
K\"all\'en-Lehmann spectral representation:
\bea
    S(q,\omega)=\frac{3\pi}{\mathcal{Z}(\beta)}\sum_{m,n}&& e^{-\beta E_m}|\langle n| S^z_q|m\rangle|^2 \times \nonumber\\
    &&\delta\left[\omega-(E_n-E_m)\right],
    \label{eq:lehmann_dyn_str_fac}
\eea
where the sum is performed over the eigenstates of the Hamiltonian~\eqref{eq:hamiltonian} with the partition function
$\mathcal{Z}(\beta)=\mathrm{Tr}\{e^{-\beta\mathcal{H}}\}$ at inverse temperature $\beta=1/k_B T$ (where we set $k_B=1$ in the following).
The momentum space operator $S_{q}^{z}$ is the Fourier transformation of the real space spin operator $S_{q}^{z}=\sum_{r}e^{-\mathrm{i}qr}S_{r}^{z}/\sqrt{L}$
with $q=2n\pi/L$ for PBCs and $S_{q}^{z}=\sum_{r}{\sqrt{2}\sin{(qr)}S_{r}^{z}}/\sqrt{L+1}$ with $q=n\pi/(L+1)$ for OBCs, where $n=1,2,\ldots,L$.
The static structure factor $S(q)$
is obtained by integrating over all frequencies,
\be
    S(q)=\frac{1}{\pi}\int_{-\infty}^{+\infty}\mathrm{d}\omega\,S(q,\omega),
    \label{eq:static_structure_factor}
\ee
which satisfies the sum rule $\sum_q S(q)=3L/4$.

While for a finite system a spectral function such as Eq.~(\ref{eq:lehmann_dyn_str_fac}) is strictly speaking a sum of $\delta$ functions, the density
of these $\delta$ functions increases rapidly with increasing system size and a smooth continuous distribution forms when some small broadening is imposed. However,
in some cases isolated $\delta$ functions with non-zero weight can remain even in the thermodynamic limit. Generically, in Eq.~\eqref{eq:lehmann_dyn_str_fac}
one can separate a $\delta$ function at $\omega=0$, which we will refer to as the singular contribution, and regular parts forming a continuum;
\be
S(q,\omega)=S_\mathrm{sing}(q,\omega) + S_\mathrm{reg}(q,\omega).
\ee
It is clear that this singular part arises from degenerate energy eigenstates ($E_m=E_n$),
\bea
    S_\mathrm{sing}(q,\omega)&=&\frac{3\pi}{\mathcal{Z}(\beta)}\sum_{m,n}^{E_m=E_n}\mathrm{e}^{-\beta E_m}|\langle n|S^z_q|m\rangle|^2\delta(\omega)\nonumber\\
    &=&a_q(\beta)\,\delta\left(\omega\right),
    \label{eq:sing_part}
\eea
with $a_q(\beta)$ the weight of the $\delta$-function at $\omega=0$. It follows that the regular part can be easily computed by constraining $E_{n}\neq E_{m}$ in
Eq.~(\ref{eq:lehmann_dyn_str_fac}), which forms a continuous distribution when some small broadening is imposed or by collecting the spectral weight in a histogram
with finite bin width.


In fact the contributions to the weight $a_q(\beta)$ of the possible singular part only can originate from eigenstates for which
$S^z_\mathrm{tot}\neq 0$ (assuming that the total number of spins is even). In the presence of disorder, there is
normally no degeneracy in the energy spectrum---apart from that related to the $\mathrm{SU}(2)$ symmetry, which does not play any role with the
operators $S^z_q$ considered here.
Accidental degeneracies can occur in principle, but would just contribute to the smooth (in the thermodynamic limit) continuum of the spectral function when $\omega\to 0$.
Disregarding these possible accidental degeneracies, the condition $E_m=E_n$ implies $|m\rangle\equiv |n\rangle$. In the sector with zero total
magnetization, we can use the spin-inversion symmetry operator $\Pi$, which flips all spins: $S^z_i \to -S^z_i$. It has the eigenvalues $p_n=\pm 1$ depending
on whether the state $|n\rangle$ has odd ($+$) or even ($-$) total spin. It is easy to show that diagonal matrix elements vanish in this sector,
\bea
    \langle n|S^z_q|n\rangle &=& \langle n|\Pi^2 S^z_q\Pi^2|n\rangle\nonumber\\
    &=& - p_n^2\langle n|S^z_q|n\rangle\nonumber\\
    &=& - \langle n|S^z_q|n\rangle,
    \label{eq:inv_symmetry}
\eea
since $\Pi S^z_q\Pi=-S^z_q$. Hence, there is no $\delta$-peak contribution from the sector of $S^z_\mathrm{tot}=0$.
For our consideration of avoiding the singular contribution, it is thus advantageous to perform thermodynamic calculations in the $S^z_\mathrm{tot}=0$ subspace
and we will refer to this as the canonical (C) ensemble --- in contrast to the grand canonical (GC) ensemble, which includes all magnetization sectors~\footnote{Using a standard Matsubara-Matsuda transformation, we can map the model onto a hardcore bosonic one, where canonical and grand-canonical are performed at fixed number of particles or chemical potential respectively.}. At zero
temperature, the C and GC ensembles are the same for most Hamiltonians since the ground state has $S^z_{\rm tot}=0$ (and also is a total-spin singlet), and
at finite temperature the two ensembles yield the same mean values for most observables in the thermodynamic limit. This last statement on the equivalence
between C and GC ensembles however does not, {\it a priori}, account for singular contributions or specific observables. For instance, evaluating the static uniform
susceptibility $\chi=d\langle S^z_{\rm tot}\rangle/dh$, where $h$ is an external magnetic field coupling to the magnetization,
\begin{equation}
\label{eq:usus}
\chi(T)=\frac{\langle(S^z_\mathrm{tot})^2\rangle}{LT},
\end{equation}
we see that it vanishes identically in the C ensemble while in the RS phase it diverges as $1/T$ when $T \to 0$ in the GC ensemble \cite{hirsch1985}. This type
of issue can, however, be avoided by considering the smallest non-zero momentum, $q=2\pi/L$, in the C ensemble, which does not correspond to a conserved
quantity.

\subsection{NMR relaxation rate}

In NMR experiments, the nuclear spins of the sample are polarized through an external magnetic field and perturbed by an electromagnetic pulse. After this
perturbation, the component of the nuclear spins along the applied magnetic field, $\mathcal{M}_z$, relaxes over time with an energy transfer to the external
environment (the lattice in a solid, or, specifically, the electrons and phonons) to reach its thermodynamic equilibrium. The demagnetization process
as a function of time $t$ can be described as
\bea
1-\mathcal{M}_z(t)\propto\mathrm{e}^{-t/T_1},
\label{mztform}
\eea
with $1/T_1$ called the spin-lattice relaxation rate~\cite{abragam1961,horvatic2002,slichter2013}. In practice, for a solid, $1/T_1$ can be directly
related to $S(q,\omega)$ according to
\bea
    \frac{1}{T_1} &=&\frac{\gamma^2}{2}\sum_q \sum_{a,b}A^2_{ab}(q) S^{ab}(q,\omega=\omega_0),
    \label{eq:nmr_def}
\eea
with $\gamma$ the gyromagnetic ratio, $A_{ab}(q)$ with $a,b=x,y,z$ the hyperfine tensor describing the coupling between nuclear and electronic spins,
and the resonance frequency $\omega_0$ given by the magnetic-field splitting between the nuclear spin levels considered, and normally one assumes $\omega_{0}\rightarrow 0$
in practice. However, in some cases spin diffusion processes can lead to significant structure even at frequencies of order typical $\omega_0$ values,
and then one needs to consider the dependence on $\omega_0$.
The hyperfine coupling is usually very short-ranged in real space. In the following, we will assume only an isotropic direct
hyperfine coupling between the nuclear spin on a given site and the electronic spin on the same site, giving a $q$-independent $A_{ab}(q)$ with only
the $a=b$ components non-zero. We further set all constants in Eq.~(\ref{eq:nmr_def}) to unity and define the spin-lattice relaxation rate of the spin chain as
\bea
\frac{1}{T_1}&=&\frac{1}{L}\sum_q S_\mathrm{reg}(q,\omega_0) =  S_\mathrm{reg}(r=0,\omega_0),
    \label{eq:nmr_shortdef_q}
\eea
where $S_\mathrm{reg}(r=0,\omega_0)$ is the regular part of the on-site dynamic structure factor. The singular $\delta$ function part of $S(q,\omega)$ should be removed (if present)
when taking the limit $\omega_0 \to 0$, which is achieved by restricting $S_{\mathrm{tot}}^{z}=0$ in our calculations. Equation~(\ref{eq:nmr_shortdef_q}) appears to be
valid only for a spatially uniform system, where the on-site spin response is independent of the lattice position. For an inhomogeneous system we compute the mean
$1/T_{1}$ averaged over all sites of the system. We will later comment on the validity of this procedure in light of the form (\ref{mztform}) of the measured spin relaxation.

%
\section{Numerical techniques}\label{sec:numerical_tech}

In this work we employ three different numerical techniques: ED, MPS and QMC-SAC, which are mostly standard or have been described in previous literature, so that we only present details sufficient for the paper to be self-contained in the Appendix.

Here we compute $S(q,\omega)$ of an open-boundary chain with $L=14$ for $D=1$ in the C ensemble as a benchmark. In ED and MPS calculations, $S(q,\omega)$ is evaluated using Eq.~(\ref{eq:lehmann_dyn_str_fac}) while in QMC simulations, it is difficult to compute $S(q,\omega)$ from the definition directly. Instead, one can reverse the relation between $S(q,\omega)$ and the imaginary-time correlation function $G_{q}(\tau)$ to obtain $S(q,\omega)$, which reads
\be
    G_{q}(\tau) = \int_{-\infty}^\infty \mathrm{d}\omega\, S(q,\omega)K(\tau,\omega),
    \label{eq:gqt_sqw}
\ee
with the kernel being $K(\tau,\omega)={\rm e}^{-\tau\omega}/\pi$ for $\omega\in(-\infty,+\infty)$.
The correlator $G_{q}(\tau)$ measures the dynamic spin-spin correlations in momentum space, where for for a given $q$ the definition is
\be
    G_{q}(\tau) = \langle {\mathbf S}_{-q}(\tau)\cdot{\mathbf S}_{q}(0)\rangle,~~~~
    \label{eq:gqt}
\ee
and ${\mathbf S}_{-q}(\tau) ={\rm e}^{\tau\mathcal{H}}\mathbf{S}_{-q}(0){\rm e}^{-\tau\mathcal{H}}$. Here we have $G_{q}(\tau)=3\braket{S^{z}_{-q}(\tau)S^{z}_{q}(0)}$. The computation of $G_{q}(\tau)$ using QMC and then its post-simulation analytic continuation are described in Sec.~\ref{app:qmc} of the Appendix.

The four panels in Fig.~\ref{fig:comparison} illustrate $S(q,\omega)$ obtained using the three different methods. One cannot expect to resolve the jagged finite-size related structure of the ED histogram with the MPS and QMC methods, but the agreement
is good with all the main profiles consisting of a narrow low-frequency peak and a broad continuum at higher frequencies. Note that there is no
strict singular part of $S(q,\omega)$ here and the low-energy peak reflects $|\omega|\not=0$ contributions distributed at very low frequencies.
We can see that the lower peak obtained with the QMC-SAC method is consistently a bit too much broadened. We will see later that this artificial
broadening effect (which in principle will go away if the error bars on the underlying imaginary-time correlation functions are reduced) diminishes
 with increasing inverse temperature $\beta$. We believe that this behavior can be traced to the fact that the effective
amount of information in the imaginary time data $G(\tau)$ increases with $\beta$, especially the long-time information, since the maximum imaginary
time is $\tau=\beta/2$.

Overall these test results give us confidence that all methods perform reasonably well.
In the following, we will use MPS and QMC to compute systems with large sizes. For the QMC calculations we will hereafter switch to the more commonly used PBCs, while keeping the OBCs with the MPS method as PBCs are not practical in that case. For large system sizes the discrepancies due to different definitions of the wavenumber $q$ will diminish.

\begin{figure}[htbp]
\includegraphics[width=\columnwidth,clip]{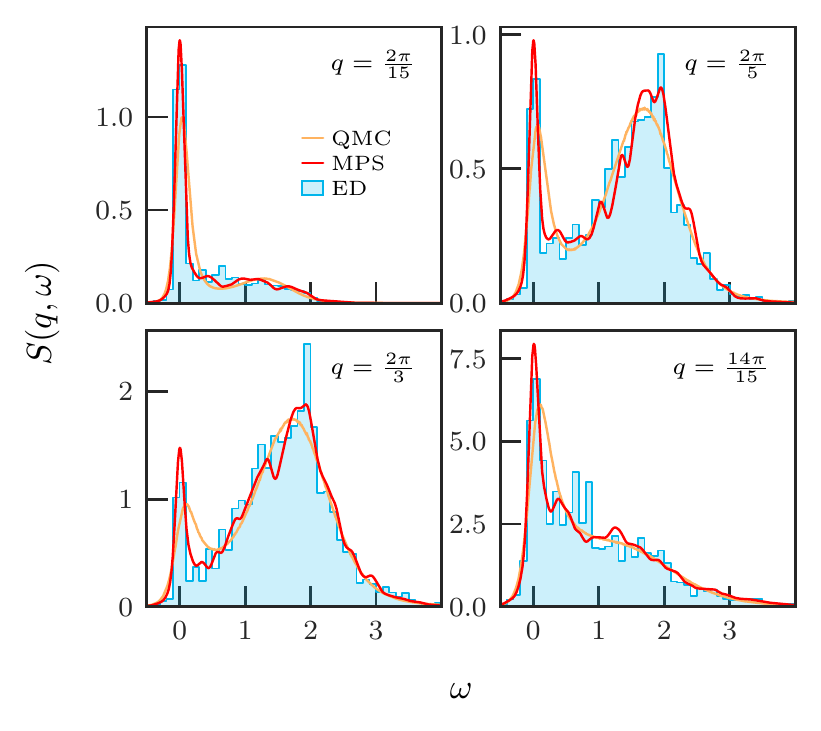}
\vskip-3mm
\caption{Dynamic structure factor at four different $q$ values at inverse temperature $\beta=8$, obtained by the three numerical methods
employed throughout the paper; ED, MPS and QMC. All calculations were performed in the canonical ensemble, on open spin chains of size $L=14$
with the couplings generated from the uniform ($D=1$) distribution. For OBCs, the wavenumber $q$ is defined as $q=n\pi/(L+1)$ with $n\in[1,L]$. The shown mean spectra represent averages over several hundred disorder
realizations.}
\label{fig:comparison}
\end{figure}

\section{Numerical results}\label{sec:results}

In this section, we first present results of $S(q,\omega)$ and discuss its scaling forms in different regimes of $q$, $\omega$,
and temperature. We then investigate the NMR relaxation rate $1/T_{1}$ extracted from the low-energy behavior of $S(q,\omega)$ and analyze the temperature
dependence of $1/T_{1}$ and its distribution. We also compute the uniform static susceptibility $\chi$ and compare it with the SDRG prediction and experimental
results, and moreover find an interesting similarity between $\chi$ and $1/T_1$. For completeness, the behavior of the static structure factor $S(q)$ is
discussed at the end of this section.
\begin{figure*}[htbp]
\includegraphics[width=2\columnwidth,clip]{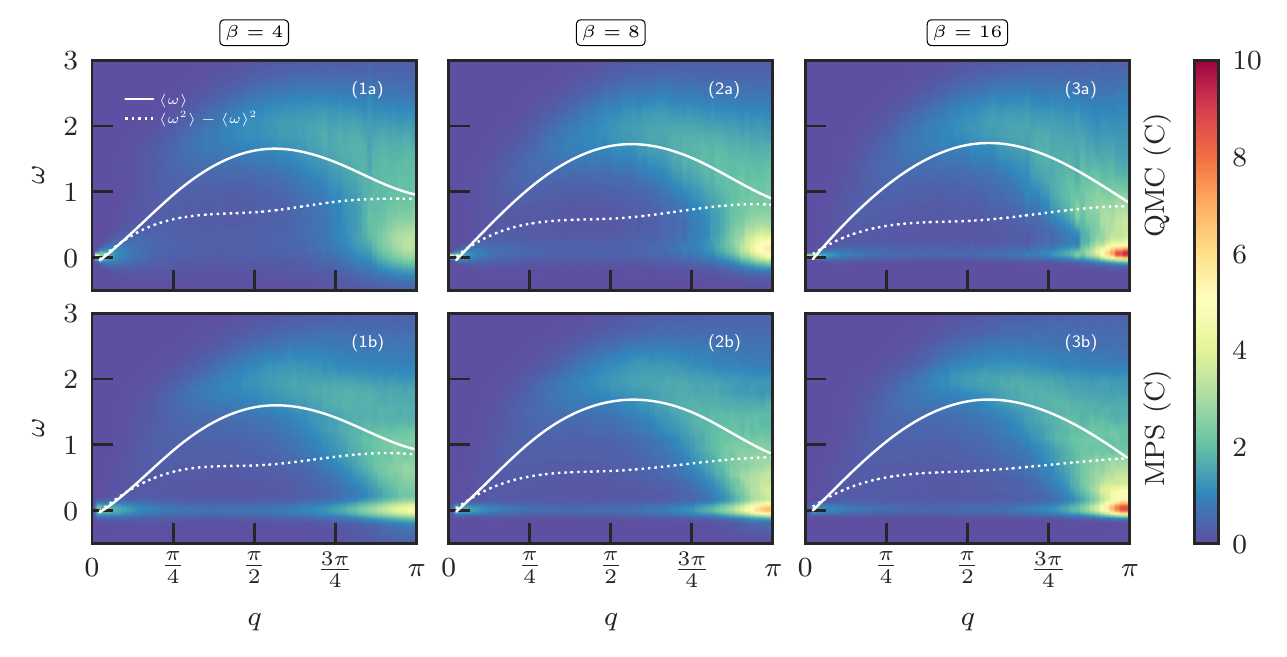}
\vskip-3mm
\caption{Comparison of QMC (upper panels) and MPS (lower panels) results for $S(q,\omega)$ at inverse temperatures $\beta=4,8$ and $16$ computed
  in the C ensemble ($S^z_\mathrm{tot}=0$). The QMC calculations were carried out on $L=128$ PBC systems and the MPS calculations on $L=64$ OBC systems,
  in both cases with the $D=1$ box distribution of couplings. The white curves show the first two cumulants, $\langle\omega\rangle$ (solid curves)
  and $\langle\omega^2\rangle-\langle\omega\rangle^2$ (dashed curves) computed directly from the spectra. We show cuts through the  $\beta=16$ data set
  at fixed frequencies and momenta in Fig.~\ref{fig:sqw_qw_cuts_D1}.}
\label{fig:sqw_eps1}
\end{figure*}

\subsection{Dynamic structure factor}
\label{subsec:sqw}

\begin{figure}[htbp]
  \includegraphics[width=8.4cm,clip]{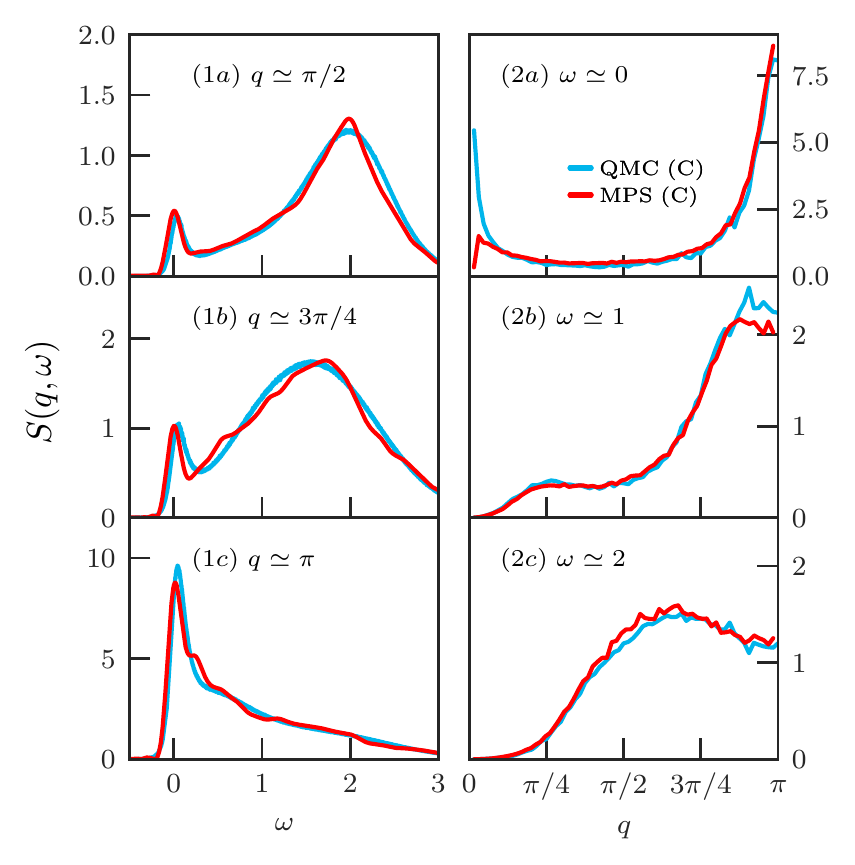}
  \vskip-3mm
\caption{Cuts at fixed momentum (left column) and fixed frequency (right column) of the dynamic structure factor $S(q,\omega)$ at $\beta=16$
  [from the same calculations as Fig.~\ref{fig:sqw_eps1}, panels ($3$a) and ($3$b)] at three different $q$ and $\omega$ values. MPS and QMC results
  are shown as blue and red curves, respectively. The slightly horizontal shift of the data at fixed momentum [panels ($1$a), ($1$b), ($1$c)] between the
  two methods is due to the different definitions of $q$ for PBC (QMC case) and OBC (MPS case). At fixed low
  energy ($\omega\simeq 0$) [panel ($2$a)], the difference at small $q$ between MPS and QMC is due to the different boundary conditions used.}
\label{fig:sqw_qw_cuts_D1}
\end{figure}

In Fig.~\ref{fig:sqw_eps1} we compare MPS and QMC results for $S(q,\omega)$ for the full range of $q$ values at three different inverse temperatures,
$\beta=4$, $8$, and $16$ for $D=1$.
In the current case (and also for the other disorder distributions, for which we do
not show the results here), the results obtained by both methods agree
very well at $\beta\simeq 16$. We make both vertical and horizontal
cuts (at fixed $q$
and $\omega$, respectively) through $S(q,\omega)$ at $\beta=16$ in Fig.~\ref{fig:sqw_qw_cuts_D1} to show the details. Only a small discrepancy can
be seen in the low $q$ modes in panel~(2a), which simply originates from the different boundary conditions (and related different definitions of $q$).
At high temperatures (smaller $\beta$), the low-energy structures are slightly broadened in the QMC results, for the reasons discussed above. As $\beta$ increases,
the imaginary-time information becomes adequate for the SAC to generate high quality spectra.
While the MPS results are clearly more reliable at high $T$, we have concluded that the QMC-SAC method is advantageous in the low-$T$ regime, where we cannot reach
sufficiently large system sizes with the MPS method. Therefore, we regard the two methods as complementary in different temperature regimes.

Comparing with the Heisenberg chain without disorder \cite{starykh1997}, we find out that the main shape of the spectra are similar, as also found
experimentally~\cite{zheludev2007} (see discussion below). However, in the disordered chains
there is also a prominent low-energy band at very low energies, with spectral weight
extending throughout the Brillouin zone but with a strong maximum around $q=\pi$ at low temperatures (and a weaker maximum around $q=0$). This feature
is a clear sign of excitations related to the high density of low-energy states in the RS state.  We observe similar results for other coupling distributions
(data not shown). For $D\simeq 0.567$, the work in Ref.~\cite{herbrych2013} found similar structures of $S(q,\omega)$, except that the low energy peaks are
claimed to be singular $\delta$ peaks, which is different from what we observe here as we have explicitly eliminated this singular contribution by working
in the C ensemble, as discussed in Sec.~\ref{sec:sing_reg_contrib}. Thus, $S(q,\omega)$ in Fig.~\ref{fig:sqw_eps1} excludes any $\delta$ peak at $\omega=0$
and only represents the regular contributions.

In the GC ensemble the amplitude of the singular peak can be determined using SAC by searching for the optimal value using the procedures developed in
Ref.~\cite{shao2017}. However, the singular peak seems to represent  a rather small fraction of the total spectral weight on average, but with large
sample-to-sample fluctuations that make an accurate determination of the mean amplitude very difficult. In the C ensemble, there should be no strict
$\delta$-function at $\omega=0$, but all our calculations still consistently show a very sharp low-frequency peak. Apart from the discrepancy regarding
the claimed singular peak, the results for $S(q,\omega)$ shown in Ref.~\cite{herbrych2013} (Fig.~S5) also appear to show a sharp structure around $q\sim\pi$
rather far above zero energy, whereas our results, for the same disorder strength and similar system size, show a more continuously evolving spectral weight
with maximal intensity at significantly lower frequency.
The reason for the different forms is not clear to us, but the consistent results from both MPS and QMC-SAC calculations in our work
(such as the near perfect agreement at $\beta=16$ in Fig.~\ref{fig:sqw_eps1}) makes us confident that these results are correct.

\begin{figure}[htbp]
    \includegraphics[width=\columnwidth,clip]{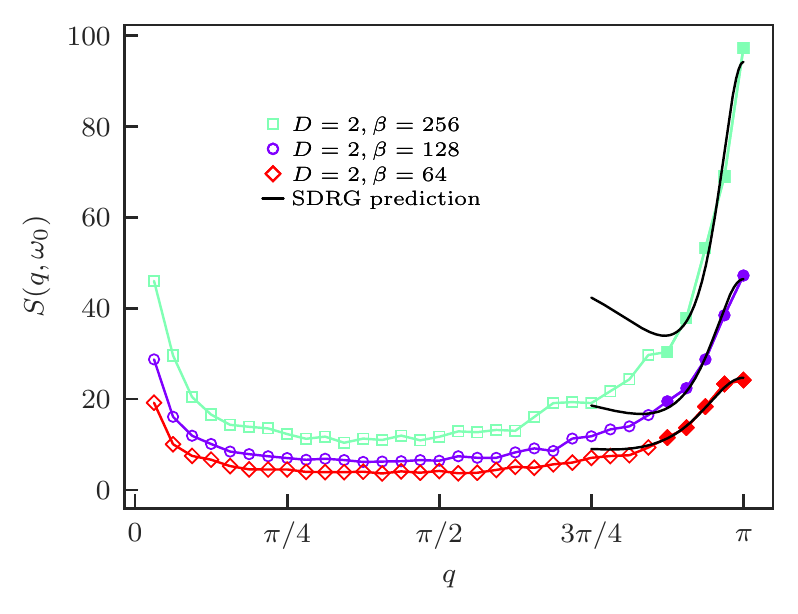}
\vskip-3mm
\caption{QMC result of $S(q,\omega)$ at fixed frequency $\omega=0.00025$ for the $D=2$ disorder distribution at three
  different temperatures; $\beta=64$, $128$, and $256$. The solid curves are fits to the SDRG scaling form $c_{1}\Phi(c_{2}\sqrt{|q-\pi|})$,
  Eq.~(\ref{eq:scaling}) at fixed $\omega$ close to $q=\pi$.  The estimated constants from these fits are $\{ c_{1},c_{2}\}=\{12.3,15.9\}$,
  $\{23.2,20.8\}$, and $\{47.1,27.0\}$, for $\beta=64$, $128$ and $256$, respectively.}
    \label{fig:scaling}
\end{figure}

Next we discuss scaling behaviors of $S(q,\omega)$ at low $T$. According to the SDRG theory, at low energies, when $q$ is close to $\pi$, at
the RS fixed point $S(q,\omega)$ obeys the scaling form~\cite{damle2000,motrunich2001}
\be
S(q,\omega)=\frac{\mathcal{A}}{l_v\omega \ln^{3}(\Omega_0/\omega)}\Phi\left[\sqrt{\left|(q-\pi)l_v\right|}\ln\left(\Omega_0/\omega\right)\right],
\label{eq:scaling}
\ee
where $\mathcal{A}$ and $l_v$ are non-universal constants and $\Omega_0$ is a cutoff energy scale. The universal scaling function $\Phi(x)$
is given by
\be
\Phi(x)=1+x\frac{\cos(x)\sinh(x)+\sin(x)\cosh(x)}{\cos^{2}(x)\sinh^{2}(x)+\sin^{2}(x)\cosh^{2}(x)}.
\label{eq:phix}
\ee
To test this form, in Fig.~\ref{fig:scaling} we plot QMC results for $S(q,\omega)$ at a fixed low frequency $\omega \approx 0$
for $L=64$, $D=2$ at three low temperatures: $\beta=64,128$, and $256$ (and we do not consider MPS calculations here because the temperatures are too
low for this method to work well). In the SDRG scheme the cutoff-scale $\Omega_0$ in Eq.~(\ref{eq:scaling}) can be interpreted as the
temperature \cite{dasgupta1980,fisher1994}. We have chosen a very small frequency $\omega=0.00025$, so that we are in the low-frequency regime
$\omega \ll \Omega_0$, as required for the scaling form to be valid \cite{damle2000,motrunich2001}, at the temperatures considered. We have fitted
the data to the form (\ref{eq:scaling}) individually for the three cases, but we can see that the trend for increasing $\beta$ ($\sim 1/\Omega_0$)
is roughly as expected based on the factor $\ln^{-3}(\Omega_0/\omega)$ which directly controls the value at $q=\pi$. We should also note here that there
may still be some finite-size effects left for the system size considered here, and in order to test the predicted form in greater detail one would have
to systematically study the convergence with increasing $L$ at fixed $\beta$ values. Nevertheless, in Fig.~\ref{fig:scaling} we observe a peak at $q=\pi$
consistent with the predicted scaling form, reflecting the predominantly AF character of the low-energy fluctuations~\cite{damle2000,motrunich2001}, but
moving away from $q=\pi$ we do not observe the minimum present in the theoretical form.
Instead we see a broader minimum at smaller $q$ (away from the region of wavenumber
where the theoretical form can be expected to apply) and a maximum as $q\to 0$ that is not predicted by the SDRG theory.

\begin{figure}[htbp]
  \includegraphics[width=\columnwidth,clip]{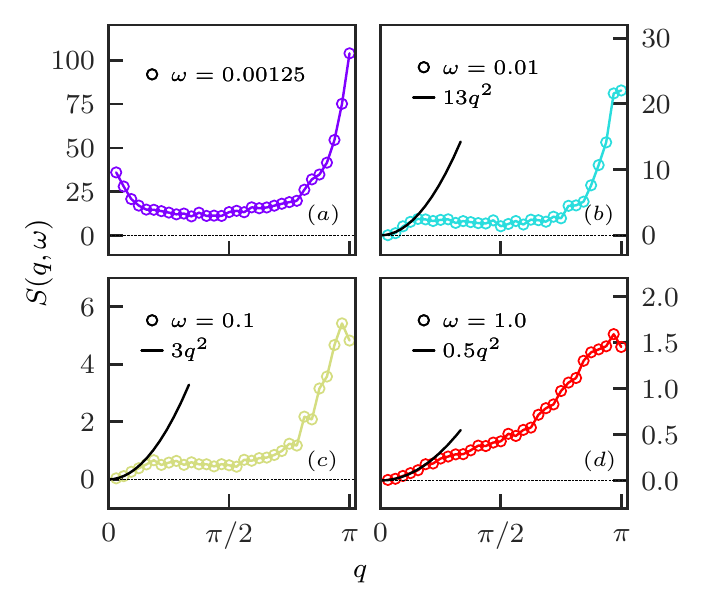}
\vskip-3mm
\caption{QMC results of $S(q,\omega)$ at different cut-off frequency $\omega$ for systems with $D=2$, $\beta=128$. Panel (a) shows that for small $\omega$,
  the contribution close to $q=0$ is divergent, while panels (b)-(d) indicate that when the horizontal cut in Fig.~\ref{fig:sqw_eps1} is made above the
  narrow band at low energy in Fig.~(\ref{fig:sqw_eps1}), the quadratic vanishing behavior predicted by the RS theory is approximately recovered.}
\label{fig:q2van}
\end{figure}

In addition to the peak structure at $q=\pi$, Ref.~\cite{motrunich2001} also predicts a quadratically vanishing behavior for small $q$;
$S(q\rightarrow 0,\omega)\propto q^{2}$ for small $\omega$, which is clearly different from our results. As the temperature is lowered, we observe a peak that
increases sharply instead. Similar behavior can be seen in
Fig.~\ref{fig:sqw_qw_cuts_D1}(2a), even though the temperature there is higher. However, as shown in Fig.~\ref{fig:q2van}, when the frequency $\omega$
of the cut is fixed at larger values, e.g., at $\omega \simeq 0.1$, a $q^{2}$ behavior is roughly reproduced. This indicates that the origin of
the low-$q$ peak at very low frequencies (likely the whole band of low-energy excitations significantly away from $q=\pi$) in Fig.~\ref{fig:sqw_eps1}
is beyond the SDRG description, but the SDRG behavior for small $q$ can still be seen approximately once one moves away from these very low
frequencies. The low-energy behavior of $S(q\sim 0,\omega\sim 0)$ is likely instead related to anomalous spin diffusion and will be important for
the NMR relaxation rate $1/T_{1}$, which we will discuss further below.

\subsection{Local spectral function}

The local spectral function $S_0(\omega)$ can be obtained directly in the MPS calculations by using the real-space spin operators $S^z_r$ instead
of the $q$ space operators in Eq.~(\ref{eq:lehmann_dyn_str_fac}). Since these calculations are done with OBCs there is some dependence on the location
$r$ within the chain even after disorder averaging has been performed, and we normally then only consider the spin at the center of the open chain.
In the QMC calculations, where we use PBCs, we can also work in real space and compute $G_0(\tau)=3\langle S^z_r(\tau)S^z_r(0)\rangle$ and apply the
SAC technique to obtain $S_0(\omega)$. This procedure can also be regarded as performing $q$ averaging of $G_q(\tau)$ before applying the SAC
method, and we will refer to it as the $q$-SAC method. Alternatively, we can average $S(q,\omega)$ over the wavenumbers after the SAC procedure has been
applied to all individual $q$ values;
\be
S_0(\omega)=\frac{1}{L}\sum_{q}S(q,\omega),
\ee
which we will refer to as SAC-$q$. In the presence of statistical noise in the QMC data, the $q$ summation and SAC procedure do not fully commute, and
$S_0(\omega)$ obtained using the different orders of operations will show some differences. In Ref.~\cite{starykh1997}, where the uniform Heisenberg chain was
studied, it was pointed out that the results are better if the $q$ averaging is done as the last step (SAC-$q$), because the functions $S(q,\omega)$ there
have a rather simple frequency profile with only a single peak, while the $q$-averaged function $S_0(\omega)$ exhibits both a sharp peak at low frequency
and a broader high-frequency maximum. The latter more complex structure is harder to resolve with analytic continuation and, therefore, the final result is
better if the $q$ averaging is done last. In the present case of the disordered chains, the individual $S(q,\omega)$ spectra also typically have two peaks,
as seen in several of the preceding figures, and therefore it is not as clear which order of operations is better.

\begin{figure}[htbp]
\includegraphics[width=\columnwidth,clip]{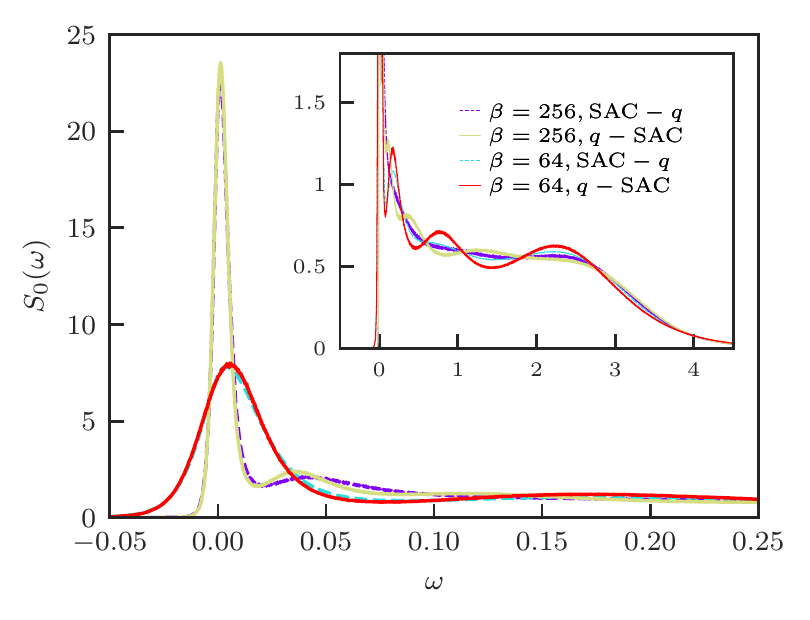}
\vskip-3mm
\caption{Comparisons at two low temperatures of the local dynamic structure factor of $L=64$ chains with $D=2$, obtained with QMC followed by the SAC technique applied before (dashed curves) and after (solid curves) the summation over $q$. The inset shows more details of the broad continuum to the right of the sharp
low-energy peak.}
\label{fig:comps0w}
\end{figure}

In Fig.~\ref{fig:comps0w} we present comparisons of results obtained with the two different orders of SAC and $q$ summation at low temperatures. Here
the two approaches deliver very similar local spectral functions, with the low-frequency peak being almost identical. Examining the details of the spectra
at higher frequencies (in the inset of Fig.~\ref{fig:comps0w}), we observe significant oscillations in the curves from the $q$-SAC method. There are some
oscillations also in the results from the SAC-$q$, but these are much smaller. It is well known that extended flat portions of a spectrum are difficult to
reproduce with analytic continuation, and most likely the large oscillations reflect a rather flat, slowly decaying $S_0(\omega)$ in the range
$\omega \sim 0.5-2.5$. The small oscillations seen in this frequency window in the SAC-$q$ should then just reflect statistical errors in the shapes
of the individual $S(q,\omega)$ spectra, which when added up still cause some fluctuations. The larger oscillations in the $q$-SAC results can
likewise be regarded as correlated statistical errors originating from the noise in $G_0(\tau)$, but with larger distortions originating from the nonlinear
way in which errors are propagated from imaginary time to real frequency through the analytic continuation procedure. Our conclusion overall is as
Ref.~\cite{starykh1997}: The structure of the individual $S(q,\omega)$ functions are less prone to distortions in analytic continuation than
the $q$-averaged spectrum $S_0(\omega)$, and, thus, it is better to apply the SAC method to the $q$-dependent data before averaging over $q$. The
fact that the dominant low-energy structure is essentially identical in the two approaches is very reassuring when considering the important
limit of small $\omega$, which enters in the spin-lattice relaxation rate that we discuss next.

\subsection{NMR relaxation rate}

\begin{figure}[htbp]
  \includegraphics[width=\columnwidth,clip]{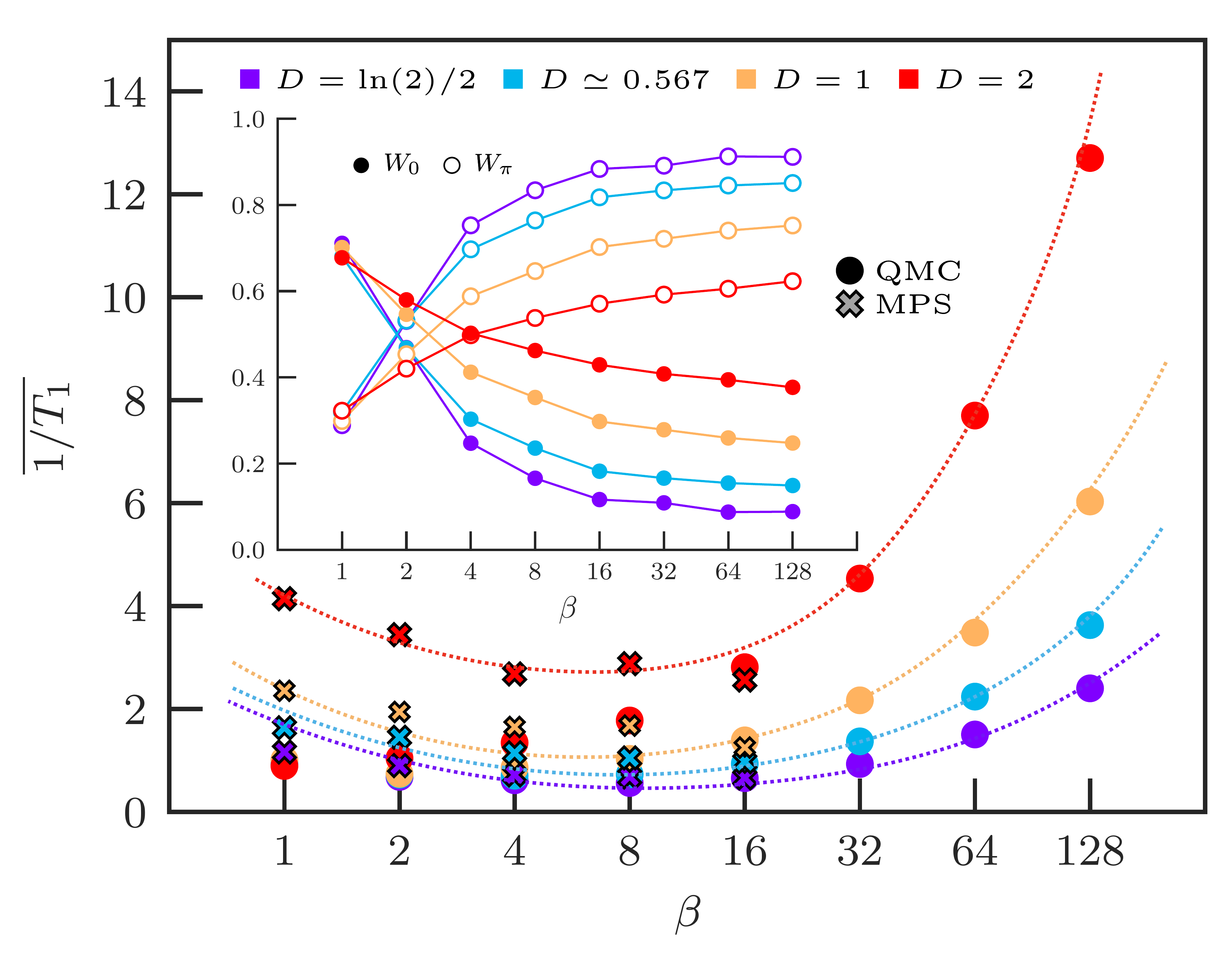}
  \vskip-3mm
    \caption{Temperature dependance of the mean NMR relaxation rate $\overline{1/T_1}$ for different disorder distributions. MPS data (shown with x symbols)
    are available only up to $\beta=16$ due to size limitations ($L=64$), while the QMC results (circles) are shown also at lower temperatures based on
    $L=128$ systems. The inset shows the relative contributions from wavevectors close to $0$ and $\pi$, as defined in Eq.~(\ref{eq:w0}). Dotted lines
    are guides to the eye.}
    \label{fig:T1_vs_beta}
\end{figure}

We have extracted the NMR spin-lattice relaxation rate $1/T_{1}$ as the mean low-energy structure factor $S_0(\omega \to 0)$ for different temperatures
and disorder distributions, using the assumption of a purely local ($q$-independent) hyperfine coupling in Eq.~(\ref{eq:nmr_def}). The results from QMC
and MPS calculations are summarized in Fig.~\ref{fig:T1_vs_beta}. In the MPS calculations
we directly averaged the local spectral function $S(r=0,\omega)$ obtained for each individual disorder realization over several hundred realizations,
while in the QMC calculations we applied the SAC-$q$ order after averaging over disorder realizations.

\subsubsection{Spin diffusion regime}

Due to the conservation of $S^{z}_{\mathrm{tot}}$, low-energy contributions are expected for small momenta $q$ in $S(q,\omega)$. This is known as
{\it spin diffusion} and is only significant at high temperature in most systems~\cite{sachdev1994,sandvik1995}. In this regime, the relaxation
rate will explicitly depends on the cutoff (length scale,  NMR frequency not being exactly zero, etc.).

In order to quantify contributions of the low $q$ and the $q\sim \pi$ modes, we separate the contributions from small $q$ and $q$ close to $\pi$.
In practice, this can be accomplished by separating the $q$ space into two equal parts \cite{sandvik1995}, defining a relative weight $W_0$ as
\begin{equation}
  W_0=\frac{\displaystyle \sum_{q\leq \pi/2}{S(q,\omega\simeq 0)}}{\sum_{q}{S(q,\omega\simeq 0)}},
  \label{eq:w0}
\end{equation}
which will typically be dominated by the $q\approx 0$ contributions. Similarly, we define $W_{\pi}=1-W_0$ to capture the $q\approx \pi$ contributions.
The temperature dependence of $W_0$ and $W_{\pi}$ are plotted in the inset of Fig.~\ref{fig:T1_vs_beta} for different disorder distributions. At high
temperatures, $W_0$ dominates $1/T_{1}$ and there exists a crossover temperature where $W_{\pi}$ overwhelms $W_{0}$. Even at the lowest temperature accessed,
$W_{0}$ is still non-negligible and grows as the disorder gets stronger, implying that spin diffusion tends to play an important role in the RS phase. This
aspect of the
dynamics appears to be beyond the prediction made within the SDRG approach (recalling that SDRG suggests vanishing behaviors of $S(q,\omega_{0})$ for low $q$ at low $T$~\cite{motrunich2001}).
It is helpful to notice that, since $W_{0}$ is normalized and evidently decays much slower than $1/T_{1}$ is growing, the un-normalized partial spectral weight
$\sum_{q\le\pi/2}{S(q,\omega\approx 0)}$ increases as $T \to 0$. This effect should be directly related to the divergence of the uniform susceptibility
$\chi$, and below we will demonstrate that both $\chi$ and $1/T_1$ indeed exhibit the same divergent behaviors.

\subsubsection{Low temperature regime}

At low temperature, we expect a more universal behavior when the spin-diffusive contributions are not important, as found, for instance, in the
unifirm chain \cite{dupont2016,coira2016}. Down to the lowest temperature accessible in the MPS calculations, $\beta\approx 16$, no tendency of increasing
$1/T_{1}$ with $\beta$ is observed for any of the disorder distributions studied. The QMC results, however, exhibit
a crossover feature from the spin diffusion regime
to the low $T$ regime, with a disorder-dependent crossover scale. As seen clearly in Fig.~\ref{fig:T1_vs_beta}, for $\beta\ge 16$, $1/T_{1}$ increases
dramatically as $T\rightarrow 0$, mainly due to the large contributions from $q$ close to $\pi$, which is similar to the uniform Heisenberg
chain \cite{sachdev1994,sandvik1995}. The contribution to $1/T_{1}$ from $q\approx 0$ is not always negligible even at low temperature,
especially for strong disorder $D$, which will be relevant for our discussion of experiments in Sec.~\ref{sec:conclusion}.
However, here we have a sharp discrepancy with both experiments and previous numerical results~\cite{shiroka2011,herbrych2013}, in which
$1/T_{1}$ was found to be decreasing as $T$ decreases. We believe such discrepancies arise from the difference between the typical and mean values of a
very broad distribution of $1/T_{1}$ values, as we will explain next.

\subsubsection{Distribution of $1/T_{1}$}

\begin{figure*}[htbp]
    \includegraphics[width=2\columnwidth,clip]{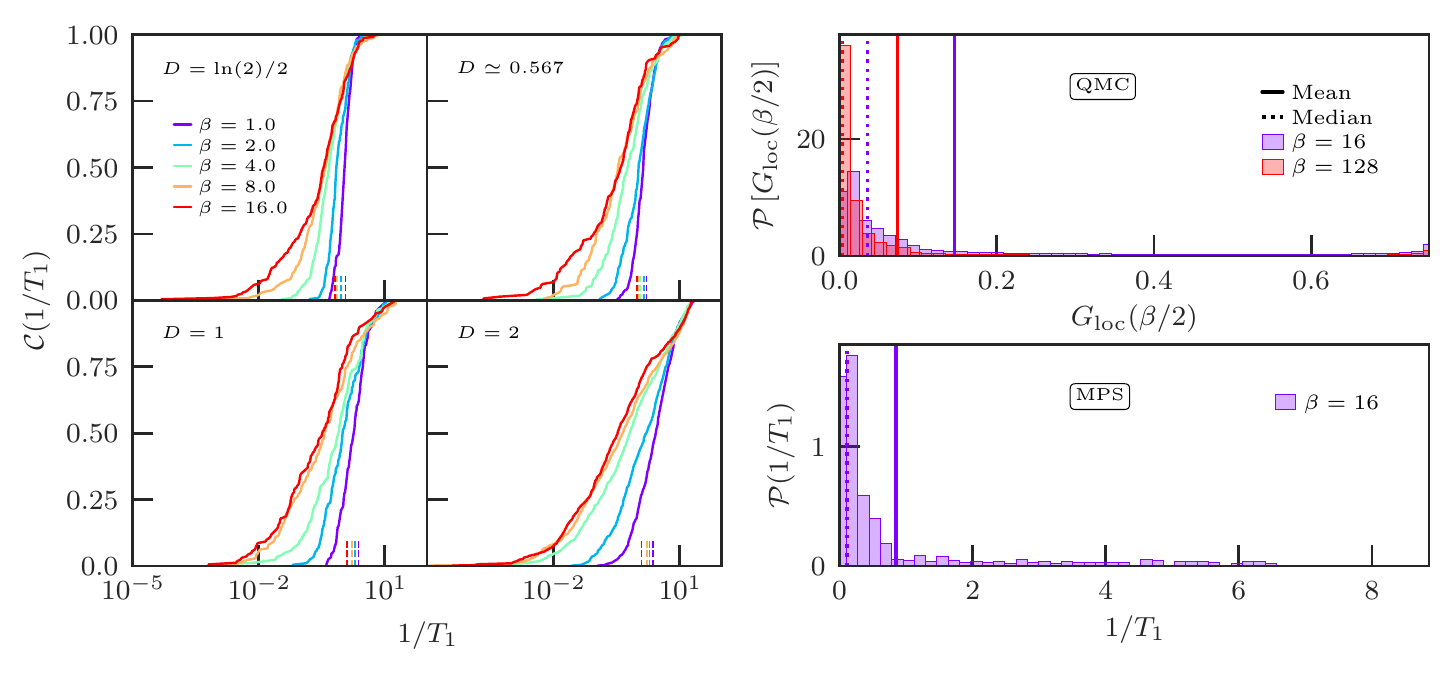}
\vskip-3mm
    \caption{Left panels: Cumulative probability distribution of the NMR relaxation rate $1/T_1$ obtained from MPS simulations for the different
      coupling distributions at $\beta=1$, $2$, $4$, $8$ and $16$. Note that the $x$ axis is shown on a logarithmic scale. The vertical marks
      indicate the corresponding mean values graphed in Fig.~\ref{fig:T1_vs_beta}. Right panels: Distributions of $1/T_1$ in $D=2$ systems
      from MPS calculations and $G_{0}(\beta/2)$ approximation from QMC calculations. The mean and median values are indicated by
      the vertical solid and dotted lines, respectively. The distributions are so broad that the mean value is dominated by the rare-event
      contribution, while the median value corresponds to typical events.}
    \label{fig:T1}
\end{figure*}

Here we consider the distribution of $1/T_{1}$. In a homogeneous
system, the locally defined
relaxation rate $1/T_1$ is also homogeneous, taking a single value, while in a disorder system one expects a distribution of relaxation rates,
associated with the measurement of the local dynamical correlation~\eqref{eq:nmr_shortdef_q} on each (inequivalent) site of the system. In such
a disordered system, the \textit{global} nuclear spin component along the applied magnetic field $\mathcal{M}_z$ is the average over each
nucleus (sites $r$) and relaxes as,
\be
    1-\mathcal{M}_z(t)\propto \frac{1}{L}\sum_r \mathrm{e}^{-t/T_{1,r}}.
    \label{eq:loc_T1}
\ee
In an experimental setup, one has only access to $\mathcal{M}_z(t)$, which can be phenomenologically modeled by a stretched exponential to take
into account disorder effects,
\be
    1-\mathcal{M}_z(t)\simeq \mathrm{e}^{-\left(t/\tau_{0}\right)^\gamma}
    \label{eq:str_exponential}
\ee
with $\tau_{0}$ and $\gamma$ fit parameters characterizing the stretched exponential distribution of $1/T_1$~\cite{johnston2006,johnston2008}.
It is evident that the uniform case should be recovered with $\gamma=1$ and $1/\tau_{0}=1/T_{1}$ at any temperature. For the disordered chains, Fig.~\ref{fig:T1}
shows the cumulative (integrated) probability distribution of $1/T_1$ computed using MPS. Note that the
calculations are carried out in real space according to Eq.~(\ref{eq:nmr_shortdef_q}), rather than in the momentum space; one cannot use the momentum space
to compute the distribution of $1/T_1$ because the Fourier transform involves an average of $1/T_1$ over each site of the chain, masking the actual
distribution of the local rates. The mean value $\overline{1/T_1}$ remains the same in the real-space
and momentum-space approaches, however. The distribution shown in Fig.~\ref{fig:T1} spreads over a few orders of magnitude and broadens as the temperature
decreases. Also, the larger the disorder strength $D$ is, the broader the distribution is at fixed temperature.

\begin{figure}[htbp]
    \includegraphics[width=\columnwidth,clip]{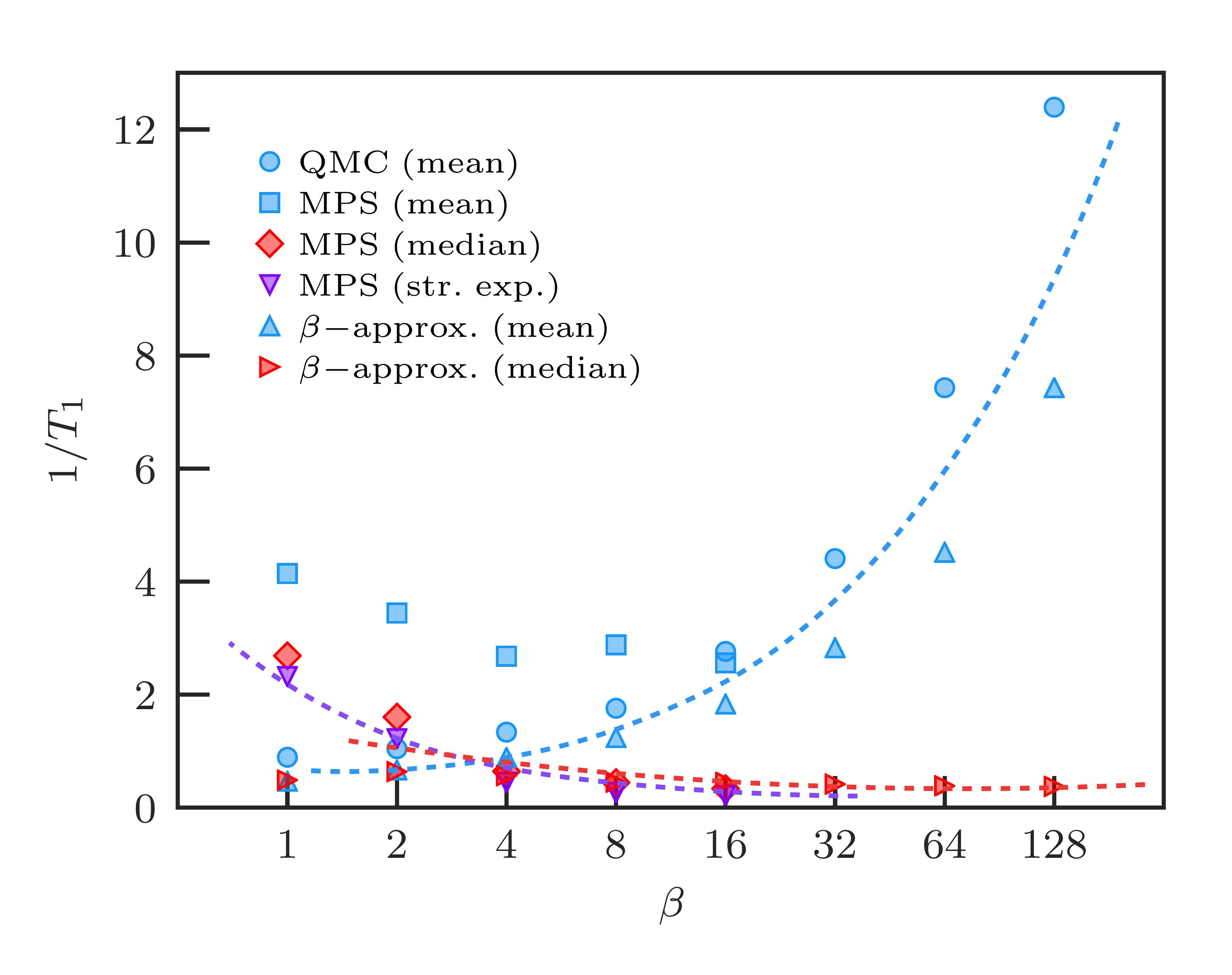}
\vskip-3mm
    \caption{Mean and median values of $1/T_1$ obtained using different methods: MPS, QMC-SAC, and QMC with $\beta$-approximation according to
      Eq.~(\ref{eq:betaapprox}). We also show the stretched exponential fit parameter $1/\tau_{0}$ defined in Eq.~\eqref{eq:str_exponential} obtained from
      the $1/T_1$ distribution of the MPS data. Symbols are the computed data point and the lines are only guides to the eye. While the mean NMR relaxation
      rate value is increasing with the inverse temperature $\beta$ (dashed blue line), the median value and the stretched exponential distributions are
      both decreasing (dashed red and purple lines).
    }
    \label{fig:t1_loc}
\end{figure}

From the distribution, the response function $\mathcal{M}_z(t)$ can be constructed and fitted to a stretched exponential as in Eq.~\eqref{eq:str_exponential},
thus determining the parameters $\tau_{0}$ and $\gamma$. The cumulative distribution in Fig.~\ref{fig:T1} implies a pronounced tail in $\mathcal{P}(1/T_{1})$,
and further, the existence of a $\gamma$ that is much smaller than $1$, especially at low $T$ and strong disorders
(data not shown)~\cite{herbrych2013}. We plot the temperature dependence of $1/\tau_{0}$ for $D=2$ in Fig.~\ref{fig:t1_loc}, and these results agree well
with previous investigations~\cite{herbrych2013,shiroka2011}. However, the stretched exponential is only an approximation,
since it neglects rare events, which can be indeed seen as some anomalously large $1/T_1$ values in Fig.~\ref{fig:T1}. These large values are very unlikely
within the stretched exponential distribution, and, thus, the resulting $1/\tau_{0}$ value only gives an estimate of a {\it typical} $1/T_1$. Here and below
we use the median value of $1/T_1$, which is relatively insensitive to the rare events, to represent a typical measurement.

In QMC calculations we do not perform the SAC procedures for individual disorder realization but always work with the disorder averaged $G(\tau)$ in order to
have sufficient statistical precision, so that we do not have access to the distribution of $1/T_1$ in this case. To study
the $1/T_1$ distribution we instead use a long-standing approximation (which we will refer here to as the $\beta$-approximation) where the $1/T_1$ value
can be be obtained \textit{directly} from imaginary-time data without full analytic continuation \cite{randeria1992}. We here provide an alternative
derivation of the $\beta$-approximation based on $1/T_1$ expressed using $S(q,\omega)$ instead of the related dynamic susceptibility $\chi(q,\omega)$
considered previously \cite{randeria1992}.

The general idea underlying the approximation is that the low-frequency behavior is most strongly reflected at the longest imaginary-time value,
$\tau=\beta/2$, accessible at inverse temperature $\beta$. At this $\tau$ point, by replacing $S(q,\omega)$ in Eq.~(\ref{eq:nmr_def}) by its Taylor expansion
around $\omega=0$ and keeping only the leading term, we obtain
\begin{eqnarray}
  \label{eq:gt2}
  G_{q}(\beta/2)&=&\frac{2}{\pi}\int_{0}^{+\infty}{d\omega e^{-\beta\omega/2}[S(q,0)+\omega S^{\prime}(q,0)+...]} \nonumber\\
  & \approx &\frac{4}{\pi\beta}S(q,0).
\end{eqnarray}
With $1/T_{1}$ defined by Eq.~(\ref{eq:nmr_shortdef_q}), the $\beta$-approximation of its value is then given by
\begin{equation}
  \label{eq:betaapprox}
  \frac{1}{T_{1}} \approx \frac{\pi}{4}\beta G_{0}(\beta/2),
\end{equation}
where the local correlation function $G_{0}(\tau)$ also equals to $\sum_{q}{G_{q}(\tau)}/L$. Similar to the MPS calculations, when considering the
distribution of the $1/T_1$ values we have to work in real space.

For the above approximation to be good, $S(q,\omega)$ should not show a strong $\omega$-dependence in the window $[0,\Delta\omega]$, where $\Delta\omega$
scales as $2/\beta$. However, in the case considered here, $S(q,\omega)$ appears to have a very high but narrow peak, with a width smaller than $2/\beta$,
in the vicinity of $\omega=0$ and a much lower, broader peak at higher energies, especially for strong disorder at low $T$. Therefore, it is reasonable
for the $\beta$-approximation to deviate from the exact results, as we indeed find. In any case, we expect the distribution of approximants to reflect the
distribution of $1/T_1$. As an aside, we note that higher-order approximants can in principle be defined by keeping more terms in the Taylor expansion in
Eq.~(\ref{eq:gt2}) and in the future we plan to develop practical procedures for accomplishing this systematically. Here we just consider the lowest-order
$\beta$-approximation.

As indicated by~Eq.~(\ref{eq:betaapprox}), at fixed temperature, $1/T_{1}$ is proportional to $G_{0}(\beta/2)$ so that $\mathcal{P}(1/T_{1})$
can be readily read off from $\mathcal{P}[G_{0}(\beta/2)]$, which we graph in Fig.~\ref{fig:T1}. We also plot MPS results for $\mathcal{P}(1/T_{1})$
at $\beta=16$, $D=2$, for comparison. The distribution $\mathcal{P}[G_0(\beta/2)]$ is bounded by the equal-time value $G_0(0)=3/4$ and
has a long tail that becomes more evident as the temperature decreases. The tail contributes large rare-event values to the mean $1/T_{1}$,
which originate from a small number of sites with almost free spins in some disorder realizations \cite{motrunich2001}.
These rare events dominate the mean value of $\mathcal{P}[G_{0}(\beta/2)]$, and this value may not be suitable
to describe the experimental NMR measurements analyzed with the assumed stretched-exponential distribution \cite{shiroka2011}. Alternatively, we use the
typical value of the distribution to represent a local probe more properly~\cite{motrunich2001}.

We presents the temperature dependence of $1/T_{1}$ for $D=2$ in Fig.~\ref{fig:t1_loc}, using the different methods discussed above. The mean values obtained
using QMC and MPS are computed in the momentum space, as also already shown Fig.~\ref{fig:T1_vs_beta}, while the other calculations were performed in real
space. Even though the $\beta$-approximation [Eq.~(\ref{eq:betaapprox})] cannot fully reproduce the QMC (mean)result, it indeed impressively catches the
expected \cite{motrunich2001} low-$T$ diverging trend. In addition, in the low-$T$ regime we observe good agreements between
Refs.~\cite{shiroka2011,herbrych2013} and the median $1/T_{1}$ extracted from the MPS results using the stretched
exponential fitting.

\subsection{Uniform susceptibility}

\begin{figure}[htbp]
  \includegraphics[width=8.3cm,clip]{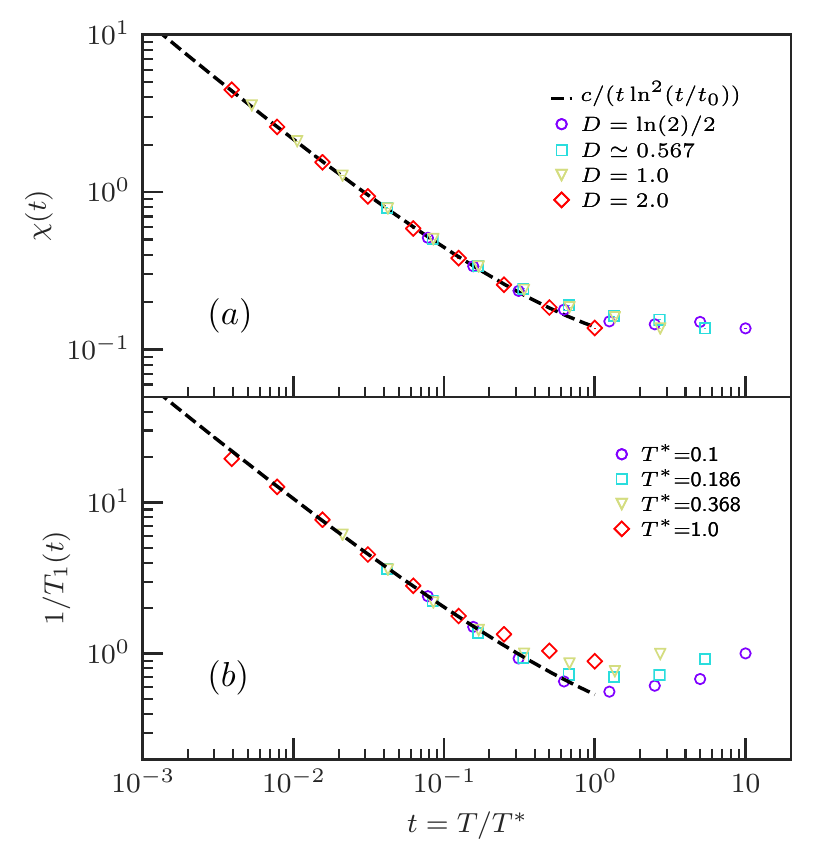}
\vskip-3mm
  \caption{(a) Temperature dependence of the uniform susceptibility for different disorders distributions obtained by QMC with PBCs. The
   temperature axis is rescaled by $T^{*}(D)$, which is chosen to be $1$ for $D=2$. The low temperature behaviors collapse onto a common
   form consistent with the SDRG prediction, Eq.~(\ref{chirsform}), indicated by the dashed curve, with parameters $c\approx 1.25$ and $t_{0}\approx 20$.
   (b) Mean spin-lattice relaxation rate rescaled using the same values of $T^{*}(D)$ as in (a). The data collapse to the same functional form, with parameters
    $c\approx 7.3$ and $t_{0}\approx 40$ in Eq.~(\ref{chirsform}).}
  \label{fig:usus}
\end{figure}

The uniform susceptibility $\chi$ is defined as Eq.~(\ref{eq:usus}) in the GC ensemble. Within the SDRG theory, its low-$T$ form is predicted to be
\be
\chi(T)= \frac{c}{T\ln^{2}(T_0/T)},
\label{chirsform}
\ee
with nonuniversal constants $c$ and $T_0$~\cite{dasgupta1980,fisher1994}.
Figure \ref{fig:usus}(a) shows the temperature dependence for different disorder distributions
obtained in QMC calculations. Here the horizontal axis is rescaled by a temperature $T^{*}$ chosen for the different cases such that the data  collapse
as much as possible onto a universal function. We arbitrarily choose $T^*=1$ for $D=2$. We find good data collapse and at low temperature the behavior is very
well described by the SDRG prediction, consistent also with previous investigations~\cite{masuda2004,shiroka2013}. Remarkably, using the same scaling
parameters $T^{*}(D)$, the mean $1/T_{1}$ value for the different disorder distributions also collapse onto a common function in the low-$T$ regime.
Furthermore, the divergent behavior of $1/T_{1}$ can be described by a same functional form as $\chi(T)$, with only a different prefactor and logarithmic scale parameter.

The divergence of $\chi(T)$ at low $T$ is directly related to the high density of low-lying excitations~\cite{fisher1994,zheludev2007},
which is also consistent with the $q$-resolved results in Fig.~\ref{fig:sqw_eps1}. It also suggests that a large low-$q$, low-$\omega$ dynamic response, i.e.,
$S(q\rightarrow 0,\omega \to 0)$, should be observed as $T\rightarrow 0$~\cite{herbrych2013}.
This would imply an increasing low-$q$ contribution to the mean $1/T_{1}$, as seen in Fig.~\ref{fig:T1_vs_beta}.
The SDRG theory predicts $S(q\rightarrow 0,\omega \to 0) \sim q^2\ln(1/\omega)/\omega$ \cite{motrunich2001}, resulting in the contribution from the low-$q$ part
to $S_{0}(\omega)$ being $\ln(1/\omega)/\omega$, and if we consider finite-temperature
as cutting off the divergence at $\omega \approx T$, we obtain the divergent form $\ln(1/T)/T$. By the same arguments, the $q\approx \pi$ contribution
to $1/T_1$ is $\ln^{-3}(1/T)/T$ according to Eqs.~(\ref{eq:scaling}) and (\ref{eq:phix}) \cite{motrunich2001}. This would appear to contradict our
findings in Fig.~\ref{fig:scaling}, where it seems that the $q=\pi$ contribution is larger and grows faster than those from $q\approx 0$, and furthermore,
the divergence of the mean $1/T_1$ in our results is slower by a factor $\ln^{-3}(1/T)$ than the SDRG prediction.
The reason for the discrepancy is not clear to us, but we can note again that we also saw clear deviations from the SDRG predictions for
small $q$ and $\omega$ in Sec.~\ref{subsec:sqw}, and logarithmic corrections to static correlation were recently found in Ref.~\cite{shu2016}. Thus,
it is plausible that not all logarithmic corrections are accounted for by the SDRG method. Given the uncertainties of the numerical analytic continuation
we can also not claim to fully resolve logarithmic corrections, though it is still intriguing that the behaviors of $\chi$ and $1/T_1$ match almost
completely in Fig.~\ref{fig:usus}. This issue should be studied further.

\subsection{Static structure factor}
\begin{figure}[htbp]
    \includegraphics[width=\columnwidth,clip]{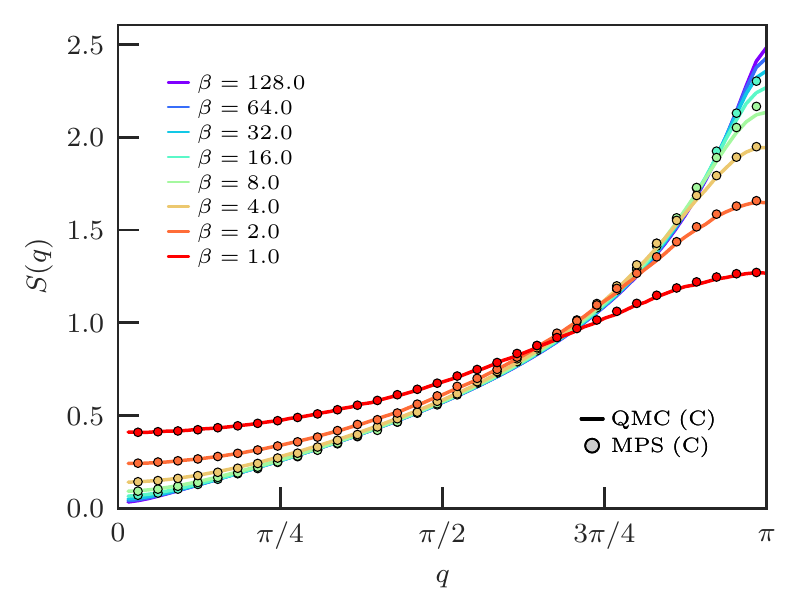}
\vskip-3mm
    \caption{Static structure factor $S(q)$ computed with QMC ($L=128$, PBCs) and MPS ($L=64$, OBCs) in the C ensemble for disorder parameter $D=1$.
    For clarity, only every second data point is displayed for the MPS data.}
    \label{fig:sq_eps1}
\end{figure}

Finally, for completeness, we discuss the frequently studied static structure factor $S(q)=G_{q}(\tau=0)$, the Fourier transform of the real-space
correlation function. In the absence of disorder, at zero temperature, the low-$q$ behavior scales linearly, $S(q\ll 1)\rightarrow K|q|/\pi$, with
$K=1/2$ being the Luttinger liquid parameter. Close to the AF wavevector, $S(q)$ diverges logarithmically $S(q)\rightarrow [-\ln(1-q/\pi)]^{3/2}$
as a result of the dominance of the spin correlation function $C(r)\sim (-1)^{r}\sqrt{\ln{(r/r_{0})}}/r$ at long distances~\cite{karbach1994,karbach1997},
where $r_{0}$ is a scale parameter. With randomness, the linear behavior at small $q$ is preserved but with a non-universal slope \cite{hoyos2007}.
In Fig.~\ref{fig:sq_eps1} we present results over a wide range of temperatures for systems with disorder parameter $D=1$. At the lowest temperature
$S(q)$ vanishes linearly as $q\to 0$ with a slope approximately $0.33$. For $q$ close to $\pi$, the divergence of $S(q)$ is completely suppressed
by randomness and the peak becomes shorter and broader~\cite{hoyos2007}, as a result of the faster decay of the mean real-space correlation function;
$C(r)\sim (-1)^{r}/r^{2}$ at $T=0$ (and a logarithmic correction to this form~\cite{shu2016}).

\section{Conclusion and discussion}
\label{sec:conclusion}
\subsection{Conclusion}
In this work, we carry out ED, MPS and QMC (supplemented by the SAC method \cite{shao2017}) calculations to study the dynamical
properties of the random $S=1/2$ Heisenberg chain.
We are able to ascertain that, in the RS phase, the finite-temperature dynamic
structure factor $S(q,\omega)$ preserves high-energy features (at $\omega$ of the order of the mean exchange constant) similar to those of the clean chain
 but broadened by the disorder, while new features appear at low energy. Most prominently, the large density of low-energy excitations, expected at low $T$ in
the RS phase, gives rise to a dispersionless narrow band of spectral weight at $\omega \approx 0$. These low-energy excitations should be
localized due to the disorder.
For $q$ close to $\pi$, $S(q,\omega)$ largely obeys the scaling form predicted by the SDRG approach~\cite{damle2000,motrunich2001}, however, the SDRG
expectation that $S(q\approx 0,\omega)\propto q^{2}$~\cite{motrunich2001} is not fulfilled in the limit of $\omega \to 0$.
Instead, we find that $S(q\approx 0,\omega \approx 0)$ is large and increases as $T$ is lowered, suggesting that, in the presence of disorder, spin diffusion
plays an important role even at very low temperatures. This diffusive feature is beyond the realm of
the SDRG method \cite{motrunich2001}.

We extracted the NMR spin-lattice relaxation rate $1/T_{1}$ from the low-energy behavior of $S(q,\omega)$ and also studied the corresponding distribution.
In the QMC-SAC approach we compute only the mean value of $1/T_1$ over the disorder, while in an experiment one can expect to probe typical values instead. To study the
distribution of $1/T_1$, we use MPS calculations as well as an approximate method \cite{randeria1992} to extract the values from the imaginary-time
QMC correlation functions that does not require full analytic continuation. 
These calculations reveal a broad distribution of $1/T_1$ values, which can be fitted using stretched exponentials, in agreement with previous investigations
~\cite{shiroka2013,herbrych2013}, except for rare very large values. The stretched exponential fitting is not sensitive to such rare contributions in the tail,
thus giving the typical value of the distribution instead of the {\it true} mean $1/T_{1}$. The typical $1/T_{1}$ exhibits a slowly decaying behavior as $T$
decreases~\cite{shiroka2013,herbrych2013}. Furthermore, we find that rare events are responsible for the divergence of the mean $1/T_1$ as $T \to 0$.
This behavior is also in agreement with analytical predictions of the excitations in the RS phase~\cite{motrunich2001}.

In addition to our main focus on dynamics, we also compute static properties, including the uniform susceptibility $\chi$ and the static
structure factor $S(q)$. The scaling form of $\chi(T)\sim 1/[T\ln^{2}(T)]$ given by the SDRG procedure \cite{fisher1994} is very well reproduced by our
results. The divergence of $\chi(T)$ ($q=0$) suggests large static and low-energy response also at small $q$, which also is consistent
with conclusion that $S(q\approx 0,\omega \approx 0)$ is large. Our results for $S(q)$ show similar linear behavior for small $q$ as seen in the clean
chain~\cite{karbach1994,karbach1997}, while for $q$ close the AF wavevector the logarithmic divergence is suppressed by randomness~\cite{hoyos2007}, in
consistent with the asymptotic $1/r^2$ decay versus distance $r$ of the real-space spin correlation function \cite{fisher1994,shu2016}.

\subsection{Relevance to experiments}

Experimentally, spin chains with random AF exchange couplings were first proposed and investigated extensively in the context of a class of
quasi-1D Bechgaard salts \cite{bulaevskii1972,azevedo1977,sanny1980,bozler1980,tippie1981,theodorou1976,theodorou1977_1,theodorou1977_2}. In these systems, a
power-law divergent magnetic susceptibility was observed, $\chi \sim 1/T^\alpha$, with $\alpha$ typically around $0.7 - 0.8$, i.e., smaller than $\alpha=1$
predicted by the SDRG theory \cite{fisher1994}, though the log correction in the theoretical result may at least be partially to blame for the
discrepancy. Another scenario is that the random Hubbard model, from which the random Heisenberg model for the materials was derived~\cite{bulaevskii1972},
has a different form of the divergence \cite{sandvik1994}. To our knowledge this lingering issue has not yet been resolved. In a more recent study of
a different material, BaCu$_2$(Si$_{0.5}$Ge$_{0.5}$)$_2$O$_7$ \cite{masuda2004,masuda2006}, the predicted RS behavior is rather well reproduced. Note
that these systems are not expected to have any ferromagnetic couplings, in which case the RS phase is not realized, due to the formation of
arbitrarily large effective magnetic moments~\cite{westerberg1994}. This situation, which we have not discussed in this paper, has been realized in
Sr$_3$CuPt$_{1-x}$Ir$_x$O$_6$ \cite{nguyen1996}.

The perhaps most extensive NMR studies of the RS state were carried on BaCu$_2$SiGeO$_7$ \cite{shiroka2013}, which can be modelled as weakly coupled
spin-1/2 Heisenberg chains with bimodal distribution of the couplings, with in-chain random couplings $J_a=24$~meV and $J_b=50$~meV~\cite{shiroka2013}, though
it is also known that neglected small three-dimensional inter-chain couplings are ultimately responsible for AF ordering at very low temperature (0.7 K)
\cite{zheludev2007} (as described, e.g., using the mean-field approach for coupled random chains \cite{yusuf2005}). In the experiments, $1/T_1$ was found to
decrease slowly as $T$ is lowered \cite{shiroka2011,shiroka2013}, as also seen in our results for the typical relaxation rate.

It should be stressed that comparisons between model calculations and experiments, especially for dynamical quantities, still have to be viewed with some caution
and further work will be needed to clarify the effects of various perturbations normally not included in the models. In fact, even for clean materials, e.g.,
BaCu$_2$Si$_2$O$_7$ (with $J=24.1$\;meV$~\simeq 280$\;K), no increase was observed in the $1/T_1$ at low-temperature in disagreement with recent numerical
studies~\cite{dupont2016,coira2016} and Luttinger-liquids predictions (where a logarithmic increase in predicted ~\cite{barzykin2000,barzykin2001}). We believe that these
discrepancies can have several explanations. First,  the small 3D coupling $J_{\mathrm{3D}}$ (which is responsible for the N\'eel ordering) in the clean
($T_N=9.2$~K) and disordered ($T_N=0.7$~K) compounds is not that small, and mean-field arguments show that the NMR relaxation can be affected already at
temperature of order $10~T_N$~\footnote{E. Orignac, private communication.}. As for the disordered system BaCu$_2$SiGeO$_7$, the NMR experiments were performed
using the $^{29}$Si nucleus~\cite{shiroka2013}, which is coupled almost symmetrically to two Cu ions (on $J_a$ bonds only), hence resulting in filtering-out
of the antisymmetric component $q\approx \pi$ by the hyperfine form factor. As a result, such NMR data would correspond to measurements primarily of the
$q\approx 0$ modes, which are responsible for a relatively small fraction of the total $1/T_1$ computed here under the assumption of a strictly on-site
hyperfine coupling.

It would be interesting to perform
NMR studies also on a different nucleus. We have not studied distributions of the low-frequency structure factor beyond the completely local on-site
correlations, and it is therefore not possible at this point to make more detailed comparisons with the experiments with a realistic hyperfine form-factor
involving also nearest-neighbor correlations. The role of mean versus typical relaxation rates in the experiments is also not fully settled, as one cannot
completely rule out that rare events also play some role. It is also clear that further work is needed to understand the role of 3D  couplings on the
RS phase and on its dynamical properties. Many of the issues pointed out above can be addressed in principle with the methods presented here.

\acknowledgments{The authors are grateful to Mladen Horvati\'c and Nicolas Laflorencie for valuable discussions and thoughtful comments. Part of this work was performed using HPC resources from GENCI (Grant No. x2016050225 and No. x2017050225) and CALMIP.
  YRS and DXY are supported by NKRDPC-2017YFA0206203, NSFC-11574404, NSFC-11275279, NSFG-2015A030313176, Special Program for Applied Research on Super Computation of the NSFC-Guangdong Joint Fund, and Leading Talent Program of Guangdong Special Projects.
  MD and SC acknowledge support of the French ANR program BOLODISS (Grant No.~ANR-14-CE32-0018), R\'egion Midi-Pyr\'en\'ees, the Condensed Matter Theory Visitors Program at Boston
  University, and ``Programme des Investissements d'Avenir'' within the ANR-11-IDEX-0002-02 program through the grant NEXT No.~ANR-10-LABX-0037.
  AWS was supported by the NSF under Grant No.~DMR-1710170 and by the Simons Foundation.}

\appendix
\section{Numerical methods}\label{app:numerical_tech}

In this appendix, we first discuss the ED method, then QMC calculations of $G(\tau)$ from which $S(q,\omega)$ can be obtained by
numerical analytic continuation. We discuss the recent version of the SAC method \cite{qin2017,shao2017} that we use for this purpose. Finally, we
describe our implementation of the MPS method to compute the structure factor.

\subsection{Exact Diagonalization}
Although limited to small system sizes, a full ED is straightforward to implement by computing all eigenstates of $\mathcal{H}$ with $S^z_{\rm tot}=0$
(C ensemble) or all $S^z_\mathrm{tot}$ sectors (GC ensemble) using standard algorithms.
Here we have applied this method for the disordered models on chains with up to $L=14$ spins.
This is already quite a demanding calculation, considering the loss of lattice symmetries in the presence of disorder and the need to average results
over hundreds of random coupling samples. Results corresponding to Eq.~\eqref{eq:lehmann_dyn_str_fac} at high temperatures can be collected in histograms
since the spectra consist of a large number of $\delta$ peaks. However, as $T$ is lowered the number of $\delta$ peaks of significant weight decreases,
and at some point the histograms do not represent well the behavior in the thermodynamic limit. For a given maximum accessible size there is therefore
some lower bound on the temperature for which the results are useful.

\begin{figure}[htbp]
    \includegraphics[width=\columnwidth,clip]{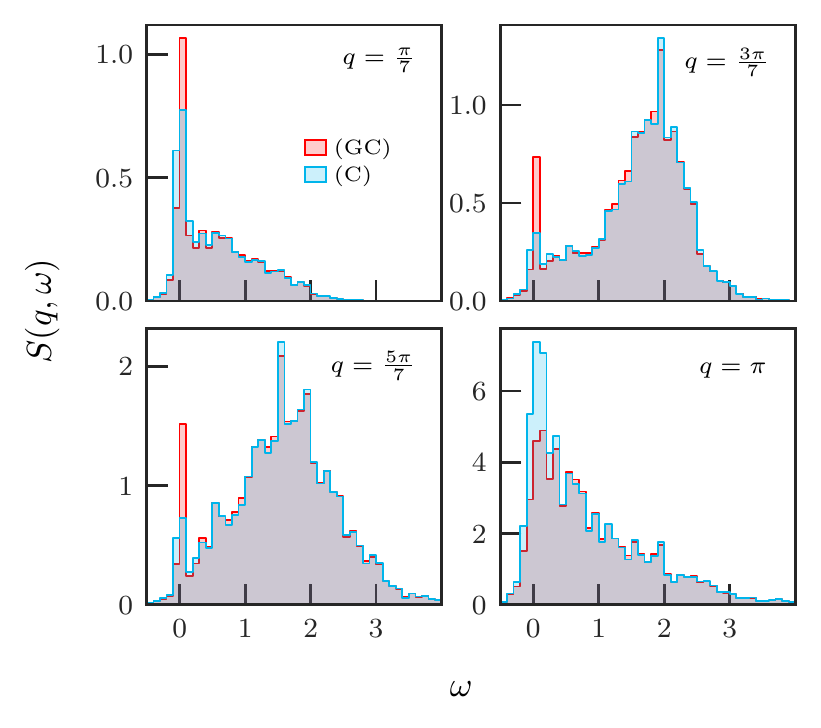}
\vskip-3mm
    \caption{ED histograms representing the dynamic structure factor for the $D=1$ model obtained with $L=14$ PBC systems at inverse temperature
      $\beta=8$. The results were averaged over 100 disorder realizations. Both C and GC results are shown. The panels correspond to different $q$
      values as indicated.}
    \label{fig:ed_gc_vs_c}
\end{figure}

ED is also quite convenient in that it allows for a simple way to compute and compare results in the C and GC ensembles. In Fig.~\ref{fig:ed_gc_vs_c},
we present typical histograms obtained for a box distributions with $D=1$ for both ensembles. The singular $\delta$-function contribution is included
in the bin centered at $\omega=0$ in the GC case. Although there is no singular contribution in the C ensemble, we still observe a significant low-energy
peak for all momenta $q$, i.e., a non-dispersive feature due to the disorder. Note also that there is no reason for the total spectral weight to be the
same in the two ensembles for small system sizes. The difference in total spectral weight is seen most clearly at $q=\pi$, where the C ensemble
produces a larger weight.

\subsection{Quantum Monte Carlo Simulations}\label{app:qmc}
\subsubsection{Stochastic Series Expansion}

In order to obtain the imaginary-time correlations, we employ the stochastic series expansion (SSE) method, which is well-known and we refer to the
literature~\cite{sandvik2010}. In this work,  we use a $\beta$-doubling trick to accelerate the equilibration stage of the simulations \cite{sandvik2002}.
The doubling process usually starts from sampling with $\beta_{0}=1$ for $N_{\mathrm{e}}$ full updating sweeps, followed by $N_{\mathrm{m}}$ sweeps
for measurements. 
The next simulation for $\beta_1=2\beta_{0}$ starts from the previous state and the doubled operator string
(two operator strings connected tail-to-tail) of the $\beta_{0}$ simulation as the initial configuration. The process is repeated until $\beta_n=\beta_02^n$ reaches the value
desired. The numbers of sweeps are typically $N_{\mathrm{e}}=1000$ and $N_{\mathrm{m}}=2N_{\mathrm{e}}$. These rather short
simulations are optimal for disorder averaging when the statistical errors are dominated by the fluctuations between different
disorder realizations, rather than the error given by the Monte Carlo statistics of an individual sample for long simulations~\cite{sandvik2002}.
The minimum length of an individual simulation is set by the time (number of updating sweeps) needed for equilibration, and the $\beta$-doubling scheme
helps to minimize this time.

\subsubsection{Stochastic Analytic Continuation}

In the SAC method, it is convenient to define another spectral function that has fixed normalization when integrated over $\omega \in [0,\infty)$.
Using the relation $S(q,-\omega)=e^{-\beta\omega}S(q,\omega)$ for $\omega > 0$, we define
\be
A_{q}(\omega)=S(q,\omega)(1+e^{-\beta\omega}),
\ee
so that
\be
    G_{q}(\tau) = \int_{0}^\infty \mathrm{d}\omega\, A(q,\omega)K(\tau,\omega),
    \label{eq:gqt_sqw2}
\ee
with the kernel in Sec.~\ref{sec:numerical_tech}, modified into
\be
    K(\tau,\omega)=\frac{1}{\pi}\frac{e^{-\tau\omega}+e^{-(\beta-\tau)\omega}}{1+e^{-\beta\omega}}.
    \label{eq:kernel2}
\ee
We can further normalize $G_q(\tau=0)$ to unity and, thus, work with a spectrum $A_{q}(\omega)$ with total spectral weight for $\omega \in [0,+\infty)$
equal to $1$. The sampling of the spectrum is performed with this weight conserved.

In our approach, $A_{q}(\omega)$ is parametrized as the sum of a large number $N_\delta$ of $\delta$ functions with fixed amplitudes $\left\{a_{n}\right\}$; normally
constant, $a_n=1/N_\delta$ \cite{shao2017}. A configuration in the SAC sampling space then represents the spectrum as
\be
    A_{q}(\omega)=\sum_{n=1}^{N_{\delta}}{a_{n}\delta(\omega-\omega_{n})},
    \label{eq:aw}
\ee
in which the positions $\left\{\omega_{n}\right\}$ of the $\delta$ functions are allowed to move within the window $(\omega_{0},\omega_{\text{m}})$, where
$\omega_{\text{m}}$ is chosen large enough for none of the $\delta$ functions to reach this limit during the sampling procedure. In our work here the spectral
weight can extend all the way to $\omega=0$ and therefore this is always taken as the lower bound (while in other cases the lower bound can be optimized
\cite{sandvik2016}). Different types of updates are carried out to sample the positions of the $\delta$ functions according to the probability
distribution
\be
    P(A_{q})\propto \exp(-\chi^{2}/2\Theta),
    \label{eq:pa}
\ee
where $\Theta$ is the sampling temperature (unrelated to the physical temperature $T$), and $\chi^{2}$ is the goodness of the fit to the QMC data,
\be
    \chi^{2}=\sum_{i=1,j}^{N_{\tau}}[\tilde{G}_{q}(\tau_{i})-G_{q}(\tau_{i})]C^{-1}_{ij}[\tilde{G}_{q}(\tau_{j})-G_{q}(\tau_{j})].
    \label{eq:chi2}
\ee
Here $\tilde{G}_{q}(\tau_{i})$ is obtained from the current spectrum $A_{q}(\omega)$ using Eq.~(\ref{eq:gqt_sqw2}) and $C$ is the covariance matrix
computed using binned data for the QMC-computed correlation function $G_{q}(\tau_{i})$.

In Eq.~(\ref{eq:pa}) $\chi^{2}$ plays the role of an energy
when treating the parametrization of $A_{q}(\omega)$ as a statistical-mechanics problem. In order to find the best averaged spectrum without over-fitting to
the noisy data, we first find the lowest possible value of $\chi^2$, denoted $\chi_{\text{min}}^{2}$, by a simulated annealing procedure starting from a
high $\Theta$ and gradually lowering it during the sampling procedure until the spectrum does not change. After this initial stage, the value of $\Theta$
is gradually increased until the averaged goodness-of-fit, $\langle \chi^{2}\rangle$, exceeds $\chi_{\text{min}}^{2}$ by an amount corresponding to the
expected standard deviation of the distribution of $\chi_{\text{min}}^{2}$, i.e.,
\be
\langle \chi^{2}\rangle \approx  \chi_{\text{min}}^{2} + \sqrt{2 \chi_{\text{min}}^{2}}.
\ee
This corresponds to a level of fluctuations at which the spectrum is properly sampled within the values of $G_q(\tau)$ corresponding to the
covariance matrix.

\begin{figure}[htbp]
\includegraphics[width=0.8\columnwidth,clip]{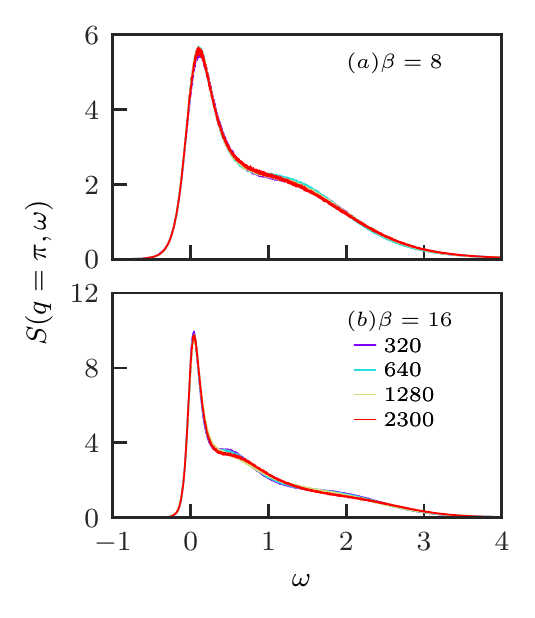}
\vskip-3mm
\caption{Convergence of $S(q=\pi,\omega)$ versus the number of disorder realizations ($320,640,1280,2300$) for two different temperatures. For $\beta=8$, the results
  converges slowly and show no trend of systematic evolution while for $\beta=16$, the spectral function changes systematically but overall, the differences are very
  small and the main structures of $S(q,\omega)$ are already steady even with a small number of disorder samples.}
\label{fig:conv}
\end{figure}

The noise in $G(\tau)$ renders the ``ill-posed'' problem of not having an unique solution in the SAC method, which in principle will be settled when the errors
are pushed to the limit of $0$. Here we study how $S(q,\omega)$ evolves as the quality of $G(\tau)$ improves. In clean systems, the quality of $G(\tau)$ is
controlled by the statistical QMC errors, while in the disordered case, the sample-to-sample variation is much larger so that the number of disorder realizations
dominates the errors of $G(\tau)$. In Fig.~\ref{fig:conv} we show the dependence of $S(q,\omega)$ on the number of disorder realizations for $\beta=8$ and $16$,
cutting at $q=\pi$ in Fig.~\ref{fig:sqw_eps1}. For $\beta=8$, the results in Fig.~\ref{fig:conv}(a) appear to be stable and do not change much as the number of
samples is increased. However, based on the comparisons with MPS calculations in the main text, we are convinced that there is still some artificial broadening
in these results. Most likely, the width of the low-frequency peak decreases so slowly with the decreasing error bars on the QMC data that it is in practice
impossible to reach the level of precision where the broadening effects become negligible. Moreover, the main features that can be resolved with reasonable
amounts of data are already seen in this case for a rather small number of disorder samples. When $\beta=16$, in Fig.~\ref{fig:conv}(b) we observe that
$S(q,\omega)$ changes systematically but very slowly, and the differences between the results from $1280$ and $2300$ samples are almost invisible. Here the
peak is much narrower than at $\beta=8$, and we know from the comparison with the MPS calculations that the width should be close to correct in this case
(and other features of the spectrum also agree very well with the MPS results).

We can explain the success at $\beta=16$ and the remaining broadening at $\beta=8$ as due to the effectively larger amount of information contained in the QMC
data at $\beta=16$ (where the range of imaginary times is twice as large). The amount of information in an imaginary-time data set depends effectively on the
range of $\tau$ values for which data are available as well as on the error bars. It appears, based on tests such as those above and in the main text,
that this effective amount of information increases quickly with increasing $\beta$, but only very slowly with decreasing error bars (and also when increasing
the number of $\tau$ points used within the available range). Thus, we conclude that, in practice, one has to live with some artificial broadening of narrow
peaks at high temperatures, while fine structures in the spectrum can be resolved increasingly well as the temperature is lowered. In order to assess the
regions where the broadening effects are not important, it is very useful to compare results of different methods, as we have done.

\subsection{Matrix Product States}

Within the MPS formalism, a wave-function can be represented as a product of local tensors $A^{p_i}_{a_ia_{i+1}}$,
\be
    |\Psi\rangle = \sum_{\{p_i\}} A^{p_1}_{a_1}A^{p_2}_{a_1a_2}\cdots A^{p_L}_{a_{L-1}}|{p_1}\rangle|p_2\rangle\cdots |p_L\rangle,
    \label{eq:general_mps}
\ee
for a chain length of $L$ sites with open boundary conditions. The local index $p_i$ is the physical index at site $i$ and can take the two values
$\uparrow$, $\downarrow$, for a spin$-1/2$. The index $a_i$ is called the virtual or bond index and encodes the entanglement of the system. Its dimension
is a control parameter, typically set to $D=500$ in this work. We use the ITensor library for these
calculations \footnote{ITensor C++ library, available at \url{http://itensor.org}}.

\subsubsection{Finite Temperature}

To address the challenge of mixed states (finite temperature), many MPS-based methods have been proposed, developed, and compared~\cite{binder2015}, e.g.,
the finite-temperature Lanczos method~\cite{kokalj2009}, the purification scheme of a pure state in an enlarged Hilbert space~\cite{verstraete2004}, and, more
recently, the minimally-entangled typically thermal states approach \cite{white2009,stoudenmire2010}. In this paper, we use the purification scheme which
introduces an auxiliary Hilbert space Q acting as a thermal bath. It can be taken as a copy of the physical Hilbert space P, thus enlarging it to
($\mathrm{P}\otimes \mathrm{Q}$) or, equivalently, doubling the system size to $2L$ with $L$ physical and $L$ auxiliary degrees of freedom. Assuming that
we know the purification of the density matrix $\rho_{\beta=0}$ as a wave-function $|\Psi_{\beta=0}\rangle$ ($\in$ P$\otimes$Q), an imaginary time evolution
performed over the infinite-temperature state will result in the desired finite-temperature state,
\begin{equation}
\ket{\Psi_\beta} = e^{-\beta\left(\mathcal{H}_P\otimes\mathcal{I}_Q\right)/2}\ket{\Psi_{\beta=0}},
\label{eq:mps_temp}
\end{equation}
with $\mathcal{H}_P$ the Hamiltonian and $\mathcal{I}_Q$ acting, respectively, on physical and auxiliary sites. The time evolution can be
carried out through the time evolution block decimation (TEBD) algorithm~\cite{vidal2004}, which relies on the Trotter decomposition of the exponential operator
in steps of size $\delta\beta=\beta/N_\mathrm{steps}$ with $N_\mathrm{steps}\gg 1$. In practice, we used a fourth order Trotter decomposition and set
$\delta\beta=0.1$. One can show that the purification of the infinite-temperature mixed state is given by the following maximally entangled state,
\begin{equation}
  |\Psi_{\beta=0}^\mathrm{GC}\rangle \propto\prod^L_{i}\Big[|\uparrow_{i,P}\downarrow_{i,Q}\rangle-|\downarrow_{i,P}\uparrow_{i,Q}\rangle\Big],
  \label{eq:mps_init_state_gc}
\end{equation}
with $\mathcal{Z}(\beta=0)=\langle\Psi_{\beta=0}|\Psi_{\beta=0}\rangle=2^L$ the partition function at infinite temperature fixing the normalization of Eq.~\eqref{eq:mps_init_state_gc}. This state corresponds to
the GC ensemble in the sense that the quantum number $S^z_\mathrm{tot}(\mathrm{P+Q})$ is fixed but not $S^z_\mathrm{tot}(\mathrm{P})$ and
$S^z_\mathrm{tot}(\mathrm{Q})$ separately. To build an initial state fulfilling that condition, one has to project~\eqref{eq:mps_init_state_gc}
onto the fixed $S^z_\mathrm{tot}(\mathrm{P})=N$ sector exclusively,
\be
    |\Psi_{\beta=0}^{\mathrm{C}(N)}\rangle=\mathcal{P}_{S^z_\mathrm{tot}(\mathrm{P})=N}|\Psi_{\beta=0}^\mathrm{GC}\rangle,
    \label{eq:mps_init_state_c}
\ee
with $\mathcal{P}_{S^z_\mathrm{tot}(\mathrm{P})=N}$ the projector operator such that the total magnetization along the $z$ axis is fixed to a given value
of $N$. This operator is hard to build as it requires the knowledge of all the ${L \choose L/2-|N|}$ states contributing to~\eqref{eq:mps_init_state_gc}.
Recent work~\cite{nocera2016,barthel2016} suggests two alternative ways to build such an initial state, which we adopt here;
\be
    \ket{\Psi_{\beta=0}^{\mathrm{C}(N)}}\propto\left(\sum_{i=1}^L S^+_{i,P}\otimes S^+_{i,Q}\right)^{\frac{L}{2}+N}\prod_{i=1}^L\ket{\downarrow_{i,P}\downarrow_{i,Q}}.
    \label{eq:mps_init_state_c2}
\ee
This state fulfills both the condition of a maximal entanglement between pairs of physical and auxiliary sites and fixed $S^z_\mathrm{tot}(\mathrm{P})$. The normalization constant in Eq.~\eqref{eq:mps_init_state_c2} is fixed by the value of the partition function at infinite temperature.

\subsubsection{Dynamic Structure Factor}

In the current MPS framework, the dynamic structure factor in Eq.~\eqref{eq:lehmann_dyn_str_fac} can be rewritten as,
\be
    S(q,\omega)=\frac{3}{\mathcal{Z}(\beta)}\langle\Psi_\beta|\left(S^z_q\otimes\mathcal{I}_Q\right)
    \delta\left(\omega-\mathcal{L}\right)\left(S^z_{-q}\otimes\mathcal{I}_Q\right)|\Psi_\beta\rangle,
    \label{eq:dy_str_fac_louv}
\ee
with $\mathcal{L}=\mathcal{H}_P\otimes\mathcal{I}_Q-\mathcal{I}_P\otimes\mathcal{H}_Q$ the Liouville operator where $\mathcal{H}_{P,Q}$ is the Hamiltonian
acting either on the physical or the auxiliary space. The eigenvalues of $\mathcal{L}$ are all the differences of the eigenenergies of the Hamiltonian,
which makes the relation with the original representation Eq.~\eqref{eq:lehmann_dyn_str_fac} more obvious. To compute the dynamical quantity in
Eq.~\eqref{eq:dy_str_fac_louv} we use the expansion in Chebyshev polynomials~\cite{holzner2011,braun2014,wolf2014,tiegel2014,tiegel2016}.

For the Chebyshev expansion to be well-defined and convergent, a first step is to map the bandwidth $W$ of the Liouvillian to $[-1, 1]$. To achieve this,
the Liouvillian operator is rescaled as
\bea
    \mathcal{L}\rightarrow\mathcal{L}'&=&\frac{1}{a}\left(\mathcal{L}+\frac{W}{2}\right)-W',\\
    \omega\rightarrow\omega'&=&\frac{1}{a}\left(\omega+\frac{W}{2}\right)-W',
    \label{eq:louv_rescaling}
\eea
with $W'=1-\varepsilon/2$, where $\varepsilon=0.025$ is a numerical safeguard to ensure that $\omega'\in[-1,1]$ and $a=W/(2W')$. The Chebyshev vectors $|t_n\rangle$
are constructed through the recursion relation
\be
    |t_n\rangle=2\mathcal{L}'|t_{n-1}\rangle-|t_{n-2}\rangle,
    \label{eq:chebyshev_recursion}
\ee
with $|t_0\rangle=(S^z_{-q}\otimes\mathcal{I}_Q)|\Psi_\beta\rangle$ and $|t_1\rangle=\mathcal{L'}|t_0\rangle$ as starting point. The Chebyshev expansion
of the dynamic structure factor~\eqref{eq:dy_str_fac_louv} up to order $N$ reads
\bea
    S(q,\omega)=&& \frac{1}{a\sqrt{1-\omega'^2}}\frac{3}{\mathcal{Z}(\beta)}\biggl[ g_0\langle t_0|t_0\rangle \nonumber\\
    && + 2\sum_{n=1}^Ng_n\langle t_0|t_n\rangle T_n(\omega')\biggr],
\eea
with $T_n(x)=\cos[n\arccos(x)]$ the Chebyshev polynomials and $g_n$ a Jackson damping factors removing oscillations due to the finite-$N$ expansion,
for which we use up to $N=3000$ for the results presented below.

Two standard alternative approaches to compute dynamical quantities with MPS would be dynamical DMRG (DDMRG)~\cite{jeckelmann2002} or real-time
evolution~\cite{vidal2004}. For a given $q$ point, the first method is quite costly as it requires a different simulation for each $\omega$ value. The
real-time evolution method allows one to compute $\langle S_{-q}^z(t)S_{q}^z(0)\rangle$ and then perform a Fourier transform to frequency space. The
latter operation can be delicate depending on the shape of the dynamical correlation versus $t$, since it can be obtained only up to a maximum
time $t_\mathrm{max}/J\sim10-20$, due to numerical limitations related to the growth of entanglement entropy with $t$. A linear
prediction technique~\cite{barthel2009} can be employed to predict the behavior of the correlator at longer times from the computed times using
a linear combination of $p$ previous data points. However, setting a systematic approach for disordered systems to find the most suitable $p$ value
is difficult in practice.

\bibliography{biblio}
\end{document}